\documentclass{elsarticle}
\usepackage{graphicx}
\usepackage{subfigure}
\usepackage{color}
\usepackage{tikz}
\usepackage{newlfont}
\usepackage{multirow}
\usepackage{array}
\usepackage{longtable} 
\usepackage{ifthen}
\usepackage{alltt}
\usepackage{float}

\usepackage{enumerate}
 \usepackage[text={6.5in,8.5in},centering]{geometry} 
\newcolumntype{C}[1]{>{\centering\let\newline\\\arraybackslash\hspace{0pt}}m{#1}}
\newcolumntype{L}[1]{>{\raggedright\let\newline\\\arraybackslash\hspace{0pt}}m{#1}}
\newcolumntype{R}[1]{>{\raggedleft\let\newline\\\arraybackslash\hspace{0pt}}m{#1}}
\newcommand{\ba}{\begin{array} }
\newcommand{\ea}{\end{array} }
\newcommand{\bae}{\begin{eqnarray}}
\newcommand{\eae}{\end{eqnarray}}
\newcommand{\bea}{\begin{eqnarray*}}
\newcommand{\eea}{\end{eqnarray*}}
\newcommand{\be}{\begin{equation}}
\newcommand{\ee}{\end{equation}}

\newcommand{\modifyr}[1]{\textcolor{red}{#1}}

\newcommand{\pr}{{\bf Proof}~~}

\usepackage{amssymb}
\usepackage{amsthm}
\usepackage{ mathrsfs }
\usepackage{amsmath}
\usepackage[mathcal]{eucal}

\usepackage{graphicx}
\usepackage{color}
\usepackage[colorlinks]{hyperref}
\usepackage{lscape}
\usepackage{graphicx}
\usepackage{epstopdf}
\newtheorem{theorem}{\hskip\parindent\bf Theorem}[section]
\newtheorem{lemma}{\hskip\parindent\bf Lemma}[section]

\newtheorem{corollary}{\hskip\parindent\bf Corollary}[section]

\usepackage{multicol}
\usepackage{tabularx}
\usepackage{ctable}
\usepackage{booktabs}
\usepackage{setspace}
\usepackage{bm}

\begin{document}
\bibliographystyle{plain}
 \markboth{Modeling of Social Insects}{Collaborations}
\title{Dynamical models of task organization in social insect colonies}
\author{Yun Kang\footnote{Sciences and Mathematics Faculty, College of Letters and Sciences, Arizona State University, Mesa, AZ 85212, USA ({\tt yun.kang@asu.edu})} and Guy Theraulaz \footnote{Centre de Recherches sur la Cognition Animale, CNRS UMR 5169, Universite Paul Sabatier, 118 route de Narbonne, 31062 Toulouse Cedex 04, France ({\tt guy.theraulaz@univ-tlse3.fr})}}
\begin{abstract}
The organizations of insect societies, such as division of labor, task allocation, collective regulation, mass action responses, have been considered as main reasons for the ecological success. In this article, we propose and study a general modeling framework that includes the following three features:  (a) the average internal response threshold for each task (the internal factor); (b) social network communications that could lead to task switching (the environmental factor); and (c) dynamical changes of task demands (the external factor). Since workers in many social insect species exhibit \emph{age polyethism},  we also extend our model to incorporate \emph{age polyethism} in which worker task preferences change with age. We apply our general modeling framework to the cases of two task groups: the inside colony task versus the outside colony task. Our analytical study of the models provides important insights and predictions on the effects of colony size, social communication, and age related task preferences on task allocation and division of labor in the adaptive dynamical environment. Our study implies that the smaller size colony invests its resource for the colony growth and allocates more workers in the risky tasks such as foraging while the larger colony shifts more workers to perform the safer tasks inside the colony. Social interactions among different task groups play an important role in shaping task allocation depending on the relative cost and demands of the tasks.\\
\end{abstract}
\bigskip
\maketitle


\section{Introduction}


{ Social insects such as ants, bees, wasps and termites, among the most diverse and ecologically important organisms on earth, live in intricately governed societies that rival our own in complexity and internal cohesion \cite{Fewell2003}. They exhibit a decentralized system of task allocation (population distribution of workers performing different tasks), with a sophisticated division of labor (different workers specializing in subsets of tasks performed by a colony) resulting from interactions among members of the colony and the environment. 
This decentralization leads to highly complex dynamics governed by many independent individual interactions, which has advantages as compared to a hierarchical organization \cite{Camazine2001, Couzin2005, Detrain2006, Detrain2008, Holbrook2009} including: \textbf{scalability} where the colony is able to adjust its organizational structure, including its division of labor and task allocations, as its size increases \cite{Holbrook2011}; \textbf{robustness/flexibility} where the colony is able to cope with environmental perturbations; and \textbf{simplicity} of the behavior of each individual in the colony. These properties have led to an increased interest in the dynamics and organization of social insect colonies in domains outside of biology, including network routing, optimization theory and robotics \cite{Dorigo1996,Pratt2009,Sumpter2010}. However, social insect biologists face the challenge of integrating between the individual and colony levels of organization \cite{Fewell2003,Wilson1971}.\\

Task allocation and division of labor are functionally inter-connected aspects of task organization.  They are fundamental properties of biological systems across all levels of organization, from cells to societies \cite{Fewell2009,MaynardSmith1995} and are two of the most prominent features of social insect colony behavior \cite{Oster1978, Wilson1968, Wilson1987}. 
Mathematical models have begun to show colony-level patterns of task organizations that can result from simple individual behavioral rules, which provide an understanding of the underlying organizational framework on which selection can act. Classical models of colony organization have focused on the adaptive value of social structure \cite{Fewell2003,Oster1978,Wilson1968,Wilson1985a,Wilson1987}. Some recent models have focused on the mechanistic processes that generate colony organization and behavior. More recent models treat the social insect colony as a self-organized, decentralized system in which behavior emerges from the independent actions and decisions of workers. Self-organization models can be used to describe numerous colony processes, including homeostasis \cite{Camazine1991,Watmough1995}, mass action responses \cite{Bonabeau1998a,Camazine1991, Camazine1999, Deneubourg1987, Deneubourg1989, Millor1999}, and colony construction \cite{Bonabeau1998a, Karsai1995}. However, modeling task organizations including division of labor and task allocations is still in an early stage. \\

 Most mechanistic models of task organization have used a simulation approach. The majority build from the assumption that individual workers vary in their response thresholds to perform a given task \cite{Hee2000,Abril2014,Porter1985,Jeanne1996,Bouwma2006}. Individuals with lower thresholds are more likely to become specialists for that task. Because individuals vary in thresholds, different group members specialize on different tasks, a division of labor emerges \cite{Bonabeau1997b,Hee2000,Porter1985,Jeanne1996,Bouwma2006,Karsai2011}.  Recent work of Gautrais et al \cite{Gautrais2002} coupled variation in response thresholds with self-reinforcement, such that individuals performing a task reduce their thresholds for performing it again. This amplifies effects of group size on specialization. Jeanson et al.\cite{Jeanson2007} generated a positive effect of group size on division of labor, from the assumption that worker availability is consistently higher than the colony's need for task performance. There is significant empirical support for these response threshold models, including demonstration that individuals have intrinsic sensory thresholds for behaviors \cite{Karsai2011,Bonabeau1998b, Page1998}, that variance in thresholds influences variation in task performance \cite{Karsai2011,Bonabeau1998b, Page1998} and that the diversity of individuals performing a task increases with stimulus levels \cite{Robinson1992,Robinson2005}. The self-reinforcement assumption also matches empirical data indicating that experienced workers are more likely to perform a task again, and to engage in it more quickly \cite{Julian1999,Beshers2001}. An age polyethism model developed by Wakano and coworkers \cite{Wakano1998} assumed that workers were grouped into different age classes, and each age class allocated labor in predetermined proportions for inside (e.g., brood care) and outside (foraging) tasks. These simulations are insightful and some of their predictions have been supported by data. However, these simulation models in many cases involve many parameters that are difficult to measure from real data; and mathematically, they are too complicated to track. In addition, researchers developing models independently often target different components of the processes generating task organizations. Thus, there is a need to broaden the integrative scope of these models by using developing and studying realistic and mathematical trackable models \cite{Fewell2003}.  \\
  
 Differential equations provide a promising tool for analyzing the mechanisms underlying colony-level patterns and dynamics such as division of labor, task allocation, and generate testable predictions about how multiple components of colony organization interact in changing environments.
These models have been successfully applied to insect societies in colony population dynamics (e.g., \cite{Kang2011-m}), colony organization (e.g., \cite{Karsai2012, Hamann2013}), foraging (e.g., \cite{Seeley1991, Sumpter2003, Udiani2015}), colony metabolic scaling (e.g., \cite{Hou2010}), optimal decision-making in social insect colonies (e.g., \cite{Marshall2009}) and social parasitism (e.g., \cite{Kang2015}). The power of these models lies in their simple mathematical formalism for describing how different interested components interact and change through time. In this manuscript, we will use compartmental differential equations models to develop a general framework to  characterize the crucial feedback mechanisms linking social interaction, colony structure and dynamics of  task organizations in a dynamical environment, and understand \emph{potential processes of task organization as colony size increases}. \\



In our proposed modeling framework, both task allocation and division of labor are modeled around the assumptions that (1) available workers perform a task when they encounter stimulus levels that match or exceed their threshold; and (2) social interaction plays the role of information transfer in task switching decisions that can affect task allocation. The later assumption has been supported by many models and data that contact rates from individuals engaged in a task increase the likelihood of recruiting others into that task. For example, in P. californicus, relative contact rates increase with worker number and density. In highly eusocial colonies (e.g., eusocial Hymenoptera \cite{Kerhoas2014, Waibel2006, Dolezal2013}), variation in worker task performance can also be generated by discrete physical castes as they age. For example, in field P. californicus colonies, younger workers are significantly less likely to perform outside tasks, including foraging and waste management, and are generally less likely to be outside the  nest \cite{Giraldo2014}. Thus, we also extend the model to include an age polyethism, in which workers shift tasks with age.  \\

 The remaining of this paper is organized as follows. In Section 2, we derive a general modeling framework that includes genetically based variation in the stimulus levels for a task to which individual workers respond; variation in task performance on differences in individual experience of the environment, task demand (flexibility) through social network communications in the colony; and the adaptive dynamics of the demand/resource of each task in the colony. In Section 3, we provide analytical results on the proposed model to address how colony size and social interaction affect the task organization. In Section 4, we extend our model to include age polyethism, and analyze the model to explore how age polyethism affects task organization in addition to the effects of colony size and social interaction. In the last section, we discuss the biological implications of our analytical results from our proposed models, and the potential outlook of our current work.\\

\section{Model derivation of task organization}

In this section, we propose  a general dynamical compartmental model of task organization on the colony level that incorporates (a) genetically based variation in the stimulus levels for a task to which individual workers respond (response thresholds \cite{Beshers2001}); 
(b) variation in task performance on differences in individual experience of the environment, task demand (flexibility) through social network communications in the colony\cite{Beshers2001, Fewell2009};
(c) dynamics of the demand/resource of each task in the colony. We assume the social insect colony has $m\geq 2$ tasks.\\

The proposed model with the three components mentioned above can be represented as the following set of nonlinear compartment model: 

\bae\label{TL-s}
\begin{array}{lcl}
N'&=& \frac{r N^s}{b+N^s}-\sum_{j=1}^m \mu_j T_j\\\\
T_i'&=&\frac{r N^s}{b+N^s} \frac{\frac{D_i}{\theta_i}}{\sum_k \frac{D_k}{\theta_k}}-\sum_j \frac{\frac{D_j}{\theta_j} T_i\frac{T_j}{N}}{\sum_k \frac{D_k}{\theta_k}} +\sum_j \frac{\frac{D_i}{\theta_i} T_j\frac{T_i}{N}}{\sum_k \frac{D_k}{\theta_k}}-\mu_iT_i\\\\
&=&\left[\sum_k \frac{D_k}{\theta_k}\right]^{-1}\left[\frac{r N^s}{b+N^s}\frac{D_i}{\theta_i}+\frac{T_i}{N}\sum_j\left(\frac{D_i}{\theta_i} -\frac{D_j}{\theta_j}\right)T_j-\mu_i T_i \sum_k \frac{D_k}{\theta_k}\right]\\\\
D_i'&=&\gamma_i N-\alpha_{i} T_i D_i
\end{array}
\eae where $\sum_i T_i'=\left(\sum_i T_i\right)'=N'.$ The variable $N$ in Model \eqref{TL-s} represents the colony size that includes all workers performing $m$ different tasks;  $T_i$ represents the size of workers performing task $i, i=1,2,..,m$ where $T(t)=(T_1(t),T_2(t),...,T_m(t))$ represents the task allocation of the colony at time $t$; and $D_i(t)$ is the work demand or stimulus level of task $i, i=1,2,..,m$. The parameter $r$ denotes the reproduction rate of queen(s) of the colony; $b$ is the saturation constant; $\theta_i$ is the average response threshold of workers performing task $i$; and $\mu_i$ is the average mortality of workers who perform task $i$.\\

 Model \eqref{TL-s} incorporates variance in response thresholds and experiences, thus division of labor and task allocation become strongly influenced by the group size $N=\sum_{i=1}^m T_i$, task number $m$, social communications among task groups $\frac{\frac{T_i}{N}\sum_j\left(\frac{D_i}{\theta_i} -\frac{D_j}{\theta_j}\right)T_j}{\sum_k \frac{D_k}{\theta_k}}, i=1,2,..,m$, and the related task demand $D_i$ where these variables showed an interaction effect \cite{Jeanson2007}. More precisely, Model \eqref{TL-s} has the following ecological assumptions:\\
\begin{enumerate}
\item The newborn workers are determined by the reproduction ability of queen(s) measured by $r$ and the collaborated efforts from $m$ task groups described by $\frac{N^s}{b+N^s}$ where $b$ is the saturation constant and $s>1$ denotes that nonlinear effects of the collaboration. This modeling approach has been used for modeling leaf-cutter ants in Kang \emph{et al} \cite{Kang2011-m} and for population dynamics of honeybees in \cite{Eberl2010,Kang2015b, Ratti2012}.\\
\item The death rate of task group $i$, denoted by $\mu_i$, should be a nonlinear function of the size of task group $i$ and the total colony size $N$. 
For simplicity, we assume $\mu_i$ is a constant and measures that average mortality of workers performing task $i$.\\
\item The demand of task $i$ of the colony, denoted by $D_i$, is determined by the following two factors: (1) The demand $\gamma_i N$ is an increasing function of the colony size where $\gamma_i$ represents the increase in demand intensity per unit time for task $i$. (2) The depletion of demand $\alpha_i T_i D_i$ is an increasing function of the demand $D_i$ and the size of task group $T_i$ where $\alpha_i$ denotes the average performance efficiency of the task group $i$. This modeling approach is adopted from Theraulaz \emph{et al} \cite{Theraulaz1998}.\\
\item The population size of task group $i$ is determined by the following two factors in addition to its natural mortality rate $\mu_i$: 
\begin{itemize}
\item The relative competition ability of task group $i$ is measured by $\frac{\frac{D_i}{\theta_i}}{\sum_k \frac{D_k}{\theta_k}}$ where $\frac{D_k}{\theta_k}$ is the competition ability of task group $k$. The term $\frac{\frac{D_i}{\theta_i}}{\sum_k \frac{D_k}{\theta_k}}$ describes that the higher demand $D_k$ and the lower response threshold  $\theta_k$ of task $k$ provide the higher competition ability of workers in task group $k$, therefore, as a consequence, $T_k$ is expected to have a higher recruitment rate of new workers.\\
\item The task switching rate between task groups is determined by their relative competition ability of task group and the task allocation at time $t$, i.e.:\\

\noindent{\textbf{The rate of worker in task group $i$ switching to other task group $j (\neq i)$:}}\\
$$f_{ij}^T= \underbrace{T_i}_\text{ size of task group $i$} \underbrace{\frac{T_j}{N}}_\text{probability to meet task group $j$} \underbrace{\frac{\frac{D_j}{\theta_j} }{\sum_k \frac{D_k}{\theta_k}}}_\text{the relative competition ability of task group $j$}$$where $ \frac{\frac{D_j}{\theta_j}\frac{T_j}{N}}{\sum_k \frac{D_k}{\theta_k}}$ describes the probability of workers performing task $i$ switching to task group $j$. Therefore, the \textbf{social network communication matrix} $\{f_{ij}^T\}_{i,j=1}^m$ contributes the task allocation due to communication such as social antennation contacts in social ants. \\
\end{itemize}

\end{enumerate}

The proposed general model \eqref{TL-s} incorporates both variation in task performance among workers and individual worker flexibility. The decision of an individual worker performing a task depends on both the internal factors (e.g., the varied thresholds for different tasks) and the external factors (e.g., task needs from the environment, or worker-worker interactions that communicate task needs). Thus, the dynamical outcomes of Model \eqref{TL-s} are expected to predict how colonies allocate workers in relation to the need for each task and adjust the allocation in response to environmental changes through the \textbf{social network communication matrix} $\{f_{ij}^T\}_{i,j=1}^m$ as colony grows. \\

For convenience, let $x_i=\frac{T_i}{N}$, then we have

$$\begin{array}{lcl}
\frac{dx_i}{dt}&=&\frac{dT_i}{Ndt}-\frac{T_idN}{N^2dt}=\frac{dT_i}{Ndt}-\frac{x_idN}{Ndt}\\\\
&=&\frac{r N^{s-1}}{b+N^s} \frac{\frac{D_i}{\theta_i}}{\sum_k \frac{D_k}{\theta_k}}-\sum_j \frac{\frac{D_j}{\theta_j} \frac{T_i}{N}\frac{T_j}{N}}{\sum_k \frac{D_k}{\theta_k}} +\sum_j \frac{\frac{D_i}{\theta_i} \frac{T_j}{N}\frac{T_i}{N}}{\sum_k \frac{D_k}{\theta_k}}-\mu_i\frac{T_i}{N}- x_i\frac{r N^{s-1}}{1+a N^s}+x_i\sum_{j=1}^m \mu_j  \frac{T_j}{N}\\\\
&=&\frac{r N^{s-1}}{b+N^s} \left[\frac{\frac{D_i}{\theta_i}}{\sum_k \frac{D_k}{\theta_k}}-x_i\right]+x_i\left[\sum_j \frac{\frac{D_i}{\theta_i}x_j}{\sum_k \frac{D_k}{\theta_k}}-\sum_j \frac{\frac{D_j}{\theta_j} x_j}{\sum_k \frac{D_k}{\theta_k}}\right] -\mu_ix_i+x_i\sum_{j=1}^m \mu_j  x_j\end{array}$$ which gives follows
$$\begin{array}{lcl}
\frac{dx_i}{dt}&=&\frac{r N^{s-1}}{b+N^s} \left[\frac{\frac{D_i}{\theta_i}}{\sum_k \frac{D_k}{\theta_k}}-x_i\right]+\frac{x_i\left[\sum_j \frac{D_i}{\theta_i}x_j-\sum_j \frac{D_j}{\theta_j} x_j\right] }{\sum_k \frac{D_k}{\theta_k}}-\mu_ix_i+x_i\sum_{j=1}^m \mu_j  x_j\\\\
&=&\frac{r N^{s-1}}{b+N^s} \left[\frac{\frac{D_i}{\theta_i}}{\sum_k \frac{D_k}{\theta_k}}-x_i\right]+\frac{x_i\left[ \frac{D_i}{\theta_i}-\sum_j \frac{D_j}{\theta_j} x_j\right] }{\sum_k \frac{D_k}{\theta_k}}-\mu_ix_i+x_i\sum_{j=1}^m \mu_j  x_j
\end{array}$$
where $x_i(0)\in [0,1] \mbox{ and } \sum_i x_i(0)=1.$ Therefore, the model \eqref{TL-s} can be rewritten as the following equations by letting $x_i=\frac{T_i}{N}, i=1,...,m$:
\bae\label{nTL-s}
\begin{array}{lcl}
N'&=& \frac{r N^s}{b+N^s}-\sum_{j=1}^m \mu_j x_j N=N\left[\frac{r N^{s-1}}{b+N^s}-\sum_{j=1}^m \mu_j x_j \right]\\\\
\frac{dx_i}{dt}&=&\frac{r N^{s-1}}{b+N^s} \left[\frac{\frac{D_i}{\theta_i}}{\sum_k \frac{D_k}{\theta_k}}-x_i\right]+\frac{x_i\left[ \frac{D_i}{\theta_i}-\sum_j \frac{D_j}{\theta_j} x_j\right] }{\sum_k \frac{D_k}{\theta_k}} -\mu_ix_i+x_i\sum_{j=1}^m \mu_j  x_j\\\\
D_i'&=&\gamma_i N-\alpha_{i} T_i D_i=\gamma_i N-\alpha_{i} x_i N D_i=N\left[\gamma_i -\alpha_{i} x_i D_i\right]
\end{array}.
\eae Notice that $x_i(0)\in [0,1],\,\, \sum_i x_i(0)=1$ and $\sum_i \frac{dx_i}{dt}=0$ for all future time $t$, thus, we can conclude that the task allocation $X=(x_1,x_2,...,x_m)$ has the properties of $x_i(t)\in [0,1],\,\,\mbox{ and } \sum_i x_i(t)=1$ for all future time $t$. In the rest of this article, we will explore the following questions by studying the dynamics of the model \eqref{nTL-s}:\\
\begin{enumerate}
\item How does the colony size $N$ affect the task allocation $X$?
\item How does the \textbf{network communication matrix} $\{f_{ij}^T\}_{i,j=1}^m$ affect the task allocation $X$?\\\\
\end{enumerate}

\section{Mathematical analysis}
Let $S=\{\sum_{i=1}^m x_i=1: 0\leq x_i\leq 1\}$. Then $\Omega=\mathbb R_+\times S\times \mathbb R_+^m$ is the state space of the model \eqref{nTL-s}. To continue our study, we first define $$f(N,\mu)=\mu(b+N^s)-rN^{s-1},\,\,N_c(\mu)=\frac{r(s-1)}{\mu s}$$  and assume that
$$\mu_1\leq\mu_2\leq\cdots\leq\mu_i\leq\cdots\leq \mu_m.$$
First, we have the following lemma:
\begin{lemma}[Existence of positive roots]\label{l1:bp} Assume $s>1$.
If $f(N_c(\mu_m),\mu_m)<0$, then there exists $N^{\pm}_1,\,\,N^{\pm}_m$ such that $f(N^{\pm}_1,\mu_1)=f(N^{\pm}_m,\mu_m)=0$, \mbox{ and } the following inequalities hold
$$N^{-}_1<N_c(\mu_1)<N^{+}_1, N^{-}_m<N_c(\mu_m)< N^{+}_m, N^{-}_1<N^{-}_m<N^{+}_m\leq N^{+}_1.$$
\end{lemma}

\noindent\textbf{Notes:} Applying Lemma \ref{l1:bp}, we could conclude that if $s=2$ and $\left(\frac{r}{\mu_m}\right)^2>4b $, then we have $$f(N_c(\mu_m),\mu_m)<0\mbox{ and  }N^{\pm}_1=\frac{r\pm\sqrt{r^2-4b\mu_1^2}}{2\mu_1}>0,\,\,N^{\pm}_m=\frac{r\pm\sqrt{r^2-4b\mu_m^2}}{2\mu_m}>0.$$ Define Condition \textbf{H} as follows:\\
$$\mbox{\textbf{H:}}\hspace{10pt} s>1,\,\,f(N_c(\mu_m),\mu_m)<0  \mbox{ and } N(0)>N^-_m.$$
Now we have our first theorem regarding the model \eqref{nTL-s} as follows:\\
\begin{theorem}[Basic dynamical properties]\label{th1:bp} Assume that all parameters are strictly positive and Condition \textbf{H} holds.
Then the model \eqref{nTL-s} is positively invariant and bounded in $\Omega$. More specifically, we have 
$$N^+_m\leq \liminf_{t\rightarrow\infty} N(t)\leq\limsup_{t\rightarrow\infty} N(t)\leq N^+_1 \mbox{ and }\frac{\gamma_i}{\alpha_i}\leq \liminf_{t\rightarrow\infty} D_i(t)\leq D_M \mbox{ for any } i=1,..,m.$$
If $N(0)<N^{-}_1$, then we have $\liminf_{t\rightarrow\infty}N(t)=0.$\\

\end{theorem}
\noindent\textbf{Notes:} Theorem \ref{th1:bp} suggests that our model \eqref{nTL-s} is well defined biologically, and the allocation of task $i$ and its related demand $D_i$ are persistent under condition that the colony is established successfully, i.e., the colony size $N$ is also persistent provided that Condition \textbf{H} is satisfied. \\

According to the model \eqref{nTL-s}, if $(N, X, D)$ is its equilibrium where $X=(x_1,...,x_m),\,\, D=(D_1,...,D_m)$, then it satisfies the following equations:
\bae\label{eq1}
\begin{array}{lcl}
0&=&\gamma_i -\alpha_{i} x_i D_i\Rightarrow D_i=\frac{\gamma_i}{\alpha_i x_i},\,\,\sum_k \frac{D_k}{\theta_k}=\sum_k \frac{\gamma_k}{\alpha_k\theta_k x_k}\mbox{ and } \sum_k \frac{D_k x_k}{\theta_k}=\sum_k \frac{\gamma_k}{\alpha_k\theta_k}\\\\
0&=&\frac{r N^{s-1}}{b+N^s}-\sum_{j=1}^m \mu_j x_j\Rightarrow \frac{r N^{s-1}}{b+N^s}=\sum_{j=1}^m \mu_j x_j\\\\
0&=&\frac{r N^{s-1}}{b+N^s} \left[\frac{\frac{D_i}{\theta_i}}{\sum_k \frac{D_k}{\theta_k}}-x_i\right]+\frac{x_i\left[ \frac{D_i}{\theta_i}-\sum_j \frac{D_j}{\theta_j} x_j\right] }{\sum_k \frac{D_k}{\theta_k}} -\mu_ix_i+x_i\sum_{j=1}^m \mu_j  x_j\end{array}\eae which give the follows
\bae\label{eq2}
\begin{array}{lcl}
0&=&\left[\sum_{j=1}^m \mu_j x_j\right] \left[\frac{\frac{D_i}{\theta_i}}{\sum_k \frac{D_k}{\theta_k}}-x_i\right]+\frac{x_i\left[ \frac{D_i}{\theta_i}-\sum_j \frac{D_j}{\theta_j} x_j\right] }{\sum_k \frac{D_k}{\theta_k}} -\mu_ix_i+x_i\sum_{j=1}^m \mu_j  x_j,\,\, \Rightarrow\\\\

0&=&\frac{\frac{D_i}{\theta_i}}{\sum_k \frac{D_k}{\theta_k}}\left[\sum_{j=1}^m \mu_j x_j+x_i\right] -\frac{x_i\left[\mu_i\sum_k \frac{D_k}{\theta_k}+\sum_j \frac{D_j}{\theta_j} x_j\right] }{\sum_k \frac{D_k}{\theta_k}} \\\\
0&=&\frac{D_i}{\theta_i}\left[\sum_{j=1}^m \mu_j x_j+x_i\right]-x_i\left[\mu_i\sum_k \frac{D_k}{\theta_k}+\sum_j \frac{D_j}{\theta_j} x_j\right] \\\\
0&=&\frac{\gamma_i}{\alpha_i\theta_i x_i}\left[\sum_{j=1}^m \mu_j x_j+x_i\right]-x_i\left[\sum_j \frac{\gamma_j}{\alpha_j\theta_j}\left(\frac{\mu_i}{x_j}+1\right)\right]
\end{array}.\eae
Therefore, we are able to solve the task allocation $X=(x_1,..,x_m)$ from the following equations
\bae\label{eqx}\frac{\gamma_i}{\alpha_i\theta_i x_i}\left[\sum_{j=1}^m \mu_j x_j+x_i\right]-x_i\left[\sum_j \frac{\gamma_j}{\alpha_j\theta_j}\left(\frac{\mu_i}{x_j}+1\right)\right]=0 \mbox{ and } \sum_{k=1}^m x_k=1,\,\,\mbox{ for any } i=1,..,m\eae which allow us to solve $D$ and $N$ from $D_i=\frac{\gamma_i}{\alpha_i x_i}$ and $\frac{r N^{s-1}}{b+N^s}=\sum_{j=1}^m \mu_j x_j$, respectively. Define $\hat{D}_i= \frac{\gamma_i}{\theta_i\alpha_i}$ as a measure of a relative demand of task $i$, then the discussion above gives the following lemma:\\

\begin{lemma}\label{l2:eq}[Equilibrium]
 If $(N^*, X^*, D^*)$ is an equilibrium of Model \eqref{nTL-s} where $X^*=(x_1^*,...,x^*_m),\,\, D^*=(D^*_1,...,D^*_m)$, then it satisfies the following equations:
 \bae\label{eqnxd}
 \begin{array}{l}
\left[\sum_{j=1}^m \mu_j x_j^*+x_i^*\right]=x^*_i\left[\sum_j \frac{\hat{D}_j}{\hat{D}_i}\left(\frac{\mu_i x^*_i}{x_j^*}+x^*_i\right)\right] \mbox{ and }\sum_{i=k}^m x_k^*=1,\,\,\mbox{ for any } i=1,..,m \\\\
D^*_i=\frac{\gamma_i}{\alpha_i x_i^*}\mbox{ and } \frac{r (N^*)^{s-1}}{b+ (N^*)^s}=\sum_{j=1}^m \mu_j x_j^*\\\\
\end{array}
 \eae
\end{lemma}
\noindent\textbf{Notes:} The solution $X$ of \eqref{eqnxd} determines by two parts:  social interactions and the recruitment rate from the relative competitive ability, i.e., the  equation
$ \begin{array}{l}\left[\sum_{j=1}^m \mu_j x_j^*+x_i^*\right]=x^*_i\left[\sum_j \frac{\hat{D}_j}{\hat{D}_i}\left(\frac{\mu_i x^*_i}{x_j^*}+x^*_i\right)\right]\end{array}$ could be represented as follows:
$$ \begin{array}{l}
\left[\left(\sum_{j=1}^m \mu_j x_j^*\right)-x^*_i\left(\sum_j \frac{\hat{D}_j}{\hat{D}_i}\frac{\mu_i x^*_i}{x_j^*}\right)\right]=\underbrace{x^*_i\left[\left(\sum_j \frac{\hat{D}_j}{\hat{D}_i}\right)x^*_i-1\right]}_\text{Contribution from social interactions}.\\
\end{array}$$

The modeling framework of the model \eqref{nTL-s} allows us investigate the following two scenarios: 
\begin{enumerate}
\item How the size of a colony $N$ affects the task allocation $X$. Since we assume that $\mu_m=\max_{1<i<m}\{\mu_i\}$ and $\mu_1=\min_{1<i<m}\{\mu_i\}$ , then According to Lemma  \ref{l2:eq}, we know that 
$$\sum_{j=1}^m \mu_j x_j^*=\mu_m-\sum_{j=1}^{m-1} (\mu_m-\mu_j) x_j^*=\mu_1+\sum_{j=2}^{m} (\mu_j-\mu_1) x_j^*.$$ Define $\Phi(N^*)=\frac{r (N^*)^{s-1}}{b+ (N^*)^s}$, then we have
\bae\label{sizeN}
\Phi(N^*)=\frac{r (N^*)^{s-1}}{b+ (N^*)^s}=\sum_{j=1}^m \mu_j x_j^*=\mu_m-\sum_{j=1}^{m-1} (\mu_m-\mu_j) x_j^*=\mu_1+\sum_{j=2}^{m} (\mu_j-\mu_1) x_j^*.
\eae
If we assume that the time scale of the population dynamics of $N$ is much slower than the time scale of the task allocation $x_i$ and the dynamics of demand $D_i$, then we could let  the total population $N^*$ be a constant when we investigate the task allocation $X$ at a faster time scale. According to \eqref{sizeN}, we can see that for a fixed population $N^*$, the value of $\Phi(N^*)$ could be increasing or decreasing with respect to $x^*_i$ depending on its corresponding mortality. We will come back to this topic when we apply our model to a two-task situation in the following subsection.

\item How social antennation interactions among task groups, i.e., the \textbf{social network communication matrix} $\{f_{ij}^T\}_{i,j=1}^m$, affects the task allocation $X$. To investigate this, we will compare the task allocation $X$ of the model \eqref{nTL-s} to the case when there is no such effects which is the following model:
\bae\label{nTL-s-nc}
\begin{array}{lcl}
N'&=& \frac{r N^s}{b+N^s}-\sum_{j=1}^m \mu_j x_j N=N\left[\frac{r N^{s-1}}{b+ N^s}-\sum_{j=1}^m \mu_j x_j \right]\\\\
\frac{dx_i}{dt}&=&\frac{r N^{s-1}}{1+a N^s} \left[\frac{\frac{D_i}{\theta_i}}{\sum_k \frac{D_k}{\theta_k}}-x_i\right] -\mu_ix_i+x_i\sum_{j=1}^m \mu_j  x_j\\\\
D_i'&=&\gamma_i N-\alpha_{i} T_i D_i=\gamma_i N-\alpha_{i} x_i N D_i=N\left[\gamma_i -\alpha_{i} x_i D_i\right]
\end{array}
\eae whose task allocation $X^*$ follows the following equation based on the similar discussions from Lemma \ref{l2:eq}:
\bae\label{eqx-nc}\begin{array}{lcl}
\sum_{j=1}^m \mu_j x_j^*&=&x^*_i\sum_j \frac{\hat{D}_j}{\hat{D}_i}\frac{\mu_i x^*_i}{x_j^*} \mbox{ and }\sum_{i=1}^m x_i^*=1, i=1,..,m\\\\
\end{array}.\eae

\end{enumerate}
\subsection{Application to the two task groups: the outside colony task versus the inside colony task\\}
To continue our study, we focus on the case when $s=2$ and $m=2$, i.e., two tasks groups (the outside colony task versus the inside colony task). It is nature to assume that the mortality of the outside colony task $\mu_2$ such as foraging  is larger than the inside colony task  $\mu_1$ such as brood care. Then $x_1$ is the task allocation for the inside colony task. This gives $x_2=1-x_1$ which is the task allocation for the outside colony task.\\

According to Lemma \ref{l2:eq} (also see equations \eqref{eq1}, \eqref{eq2}, and \eqref{eqx}), we have the following equations of the interior equilibrium $(N, X, D)$ of Model \eqref{nTL-s} when $s=2$ and $m=2$:
$$\frac{r N}{b+ N^2}= \mu_1 x_1+\mu_2 (1-x_1),\,\,D_1=\frac{\gamma_1}{\alpha_1x_1},\,D_2=\frac{\gamma_2}{\alpha_2(1-x_1)},$$ 
and $\left[1-x_1\right]\left[(1+\mu_1-\mu_2) x_1+\mu_2 \right]=x_1\left[\mu_1\left( 1+\left(\frac{\hat{D}_2}{\hat{D}_1}-1\right)x_1\right)+x_1(1-x_1)\left( 1+\frac{\hat{D}_2}{\hat{D}_1}\right)\right]$ which gives
\bae\label{eqx2}
\begin{array}{lcl}\left[1-x_1\right]^2&=&\frac{\mu_1\hat{D}_2}{\mu_2\hat{D}_1}x_1^2+\underbrace{\frac{x_1(1-x_1)}{\mu_2}\left[\left( 1+\frac{\hat{D}_2}{\hat{D}_1}\right)x_1-1\right]}_\text{contribution from social interactions}
\end{array}.\eae
Now we have the following theorem:\\

\begin{theorem}[Task allocation]\label{th2:allocation} Let $s=m=2,\,\,\mu_1<\mu_2$, and  assume that Condition \textbf{H} holds. Then the model \eqref{nTL-s} has two interior equilibria $(N^*_\pm,X^*,D^*)=\left(\frac{r\pm\sqrt{r^2-4b\mu^2}}{2\mu}, x_1^*,1-x_1^*, \frac{\gamma_1}{x_1^*\alpha_1} , \frac{\gamma_2}{(1-x_1^*)\alpha_2}\right)$ where $\mu=\mu_1x_1^*+\mu_2(1-x_1^*)$, and $x_1^*\in(0,1)$ which is the positive root of \eqref{eqx2}. Moreover, the following statements hold:
\begin{enumerate}
\item The equilibrium $(N^*_-,X^*,D^*)$ is unstable,  and the population $N^*_\pm$ are in the following intervals:
$$N^*_+\in\left(\frac{r+\sqrt{r^2-4b\mu^2_2}}{2\mu_2},\frac{r+\sqrt{r^2-4b\mu^2_1}}{2\mu_1}\right)\mbox{ and } N^*_-\in\left(\frac{r-\sqrt{r^2-4b\mu^2_1}}{2\mu_1},\frac{r-\sqrt{r^2-4b\mu^2_2}}{2\mu_2}\right).$$ 
In addition, $N^*_+$ is  an increasing function of $x_1^*$, while it is a decreasing function of $x_2^*, \mu_1, \frac{\hat{D}_2}{\hat{D}_1}$, respectively. 
\item The allocation of the inside colony task $x_1^*$ is greater than the critical value $\hat{x}_1$ where
$$\hat{x}_1=\frac{\frac{\hat{D}_2}{\hat{D}_1}(1+\mu_1)+2-\sqrt{\frac{\hat{D}_2}{\hat{D}_1}^2(1+\mu_1)^2+4\frac{\mu_1\hat{D}_2}{\hat{D}_1}}}{2(1+\frac{\hat{D}_2}{\hat{D}_1})}<\frac{1}{1+\frac{\hat{D}_2}{\hat{D}_1}}\in (0,1).$$
\item The allocation of the inside colony task $x_1^*$ is increasing with respect to $\frac{\hat{D}_1}{\hat{D}_2}, \mu_2$, respectively, and  $x_1^*$ is decreasing with respect to its average mortality rate $\mu_1$.\\
\end{enumerate}
\end{theorem}

\noindent\textbf{Notes:} Theorem \ref{th2:allocation} provides important insights on the long-term dynamics (i.e., matured) of the task organization in social insect colonies, provided that Condition \textbf{H} holds. Both theoretical results and numerical simulations (not show here) suggest that the model \eqref{nTL-s} has two interior equilibria $(N^*_\pm,X^*,D^*)$ where $(N^*_-,X^*,D^*)$ is a saddle and  $(N^*_+,X^*,D^*)$ is a sink. There is a threshold $N_c$ between $N^*_-$ and $N^*_+$ where, biologically, the smaller population size  $N$ (e.g., $N\in (N^*_-, N_c)$) indicates that the colony is at the early ergonomic growth stage which is highly unstable (see Kang \emph{et al} \cite{Kang2011-m} and Clark and Fewell \cite{Clark2014}).  When the colony size is larger than the threshold $N_c$, then the colony size approaches to the matured size $N^*_+=\frac{r+\sqrt{r^2-4b\mu^2}}{2\mu}$ where $\mu=\mu_1x_1^*+\mu_2(1-x_1^*)=\mu_2-(\mu_2-\mu_1)x_1^*=\mu_1+(\mu_2-\mu_1)x_2^*$.  According to Theorem \ref{th2:allocation}, 
$N^*_+$ is decreasing with respect to the allocation of the outside colony task $x_2^*$ and is increasing with respect to the allocation of the inside colony task $x_1^*$. This result suggests that \textbf{the matured colony} distributes more allocation to the inside colony task distribution $x_1^*$ (e.g., invests more in less risky task) as its size increases. Theorem \ref{th2:allocation} also implies that the average mortality and the related colony demand can have huge influences on a social insect colony size since  the population size of \textbf{the matured colony} is decreasing with respect to the mortality of the  inside colony task $\mu_1$ and the relative demand of the inside colony task $\frac{\hat{D}_2}{\hat{D}_1}$. More detailed discussion on the effects of the colony size will be presented in the next subsection. \\

In the case that Condition \textbf{H} does not hold,  the model \eqref{nTL-s}  can still have the stable equilibrium $(N^*_+,X^*,D^*)$ under restriction that $x_1^*$ should be greater than certain threshold $x^c_1$. The existence of $N^*_+$ requires that $\mu_1<\frac{r}{2\sqrt{b}}$ and $\mu=\mu_1x_1^*+\mu_2(1-x_1^*)<\frac{r}{2\sqrt{b}}$ which implies that $1>x_1^*>x^c_1=\frac{\frac{r}{2\sqrt{b}}-\mu_2}{\mu_1-\mu_2}$ where $x_1^*$ is the unique positive root of the equation \eqref{eqx2}. In the case that $x_1^*<x_1^c$ which is equivalent to $x^*_2>1-x_1^c$, the colony collapses. This implies that if the mortality of the outside colony task such as foraging is too high, then the colony should restrict the allocation to this task under certain threshold for survival while the allocation of the inside colony task should be above certain threshold to maintain the colony.  \\

In addition, Theorem \ref{th2:allocation} provides insights on how the relative demand and the mortality of tasks would affect the task allocation. The theorem implies that the inside colony task allocation $x_1^*$ should be above a threshold $\hat{x}_1=\frac{\frac{\hat{D}_2}{\hat{D}_1}(1+\mu_1)+2-\sqrt{\frac{\hat{D}_2}{\hat{D}_1}^2(1+\mu_1)^2+4\frac{\mu_1\hat{D}_2}{\hat{D}_1}}}{2(1+\frac{\hat{D}_2}{\hat{D}_1})}$ which is determined by its task mortality $\mu_1$ and the relative task demand $\frac{\hat{D}_2}{\hat{D}_1}$.  The threshold $\hat{x}_1$ is decreasing with respect to $\frac{\hat{D}_2}{\hat{D}_1}$ and $\mu_1$, respectively. This suggests that increasing the values of $\frac{\hat{D}_2}{\hat{D}_1}$ or $\mu_1$ could potentially decrease the inside task allocation $x_1^*$ as the task allocation is closely regulated by the colony demand  and its mortality. This confirms our second results of Theorem \ref{th2:allocation}, i.e., the allocation of the inside colony task $x_1^*$ is increasing with respect to $\frac{\hat{D}_1}{\hat{D}_2}, \mu_2$, respectively, however $x_1^*$ is decreasing with respect to $\mu_1$. These analytical results could have the following profound biological implications:
\begin{enumerate}

\item When the value of $\frac{\hat{D}_2}{\hat{D}_1}$ increases, this implies that the colony has higher demand of the outside task such as foraging, thus the colony would distribute more workers to the outside colony task $x_2^*=1-x_1^*$ despite that this task has a higher mortality. The consequence is that the colony would decrease the allocation to the inside colony task $x_1^*$

\item When the value of the average mortality rate of the inside colony task $\mu_1$ increases, the colony experiences more cost to distribute workers to this task. To optimize the investment of the colony, the colony would decrease the distribution to the inside colony task.

\item When the value of the average mortality rate of the outside colony task $\mu_2$ increases,  the colony experiences more cost to distribute workers to the outside colony task, thus the colony will increase the worker distribution to the inside colony task to optimize the energy investment. \\
\end{enumerate}


\subsection{Effects of the colony size $N$\\}  

In the previous subsection, under the assumption that population dynamics $N$ occur at the comparable time scale of the task allocation, the results of Theorem \ref{th2:allocation} provide important insights on the relationship between the long-term task allocation dynamics and the long-term population dynamics when the colony is matured (i.e., the population is above certain threshold). In this subsection, we assume that the time scale of the population dynamics of $N$ is much slower than the time scale of the task allocation $x_i$ and the dynamics of the demand $D_i$. Thus, we could let  the total population $N$ be a constant.  According to \eqref{sizeN}, we can conclude that the task allocation $X^*$ satisfies the following equations:

\bae\label{eq-constant-N}
\begin{array}{lcl}
\Phi(N^*)=\frac{r (N^*)^{s-1}}{b+ (N^*)^s}=\mu_2- (\mu_2-\mu_1)x_1^*=\mu_1+(\mu_2-\mu_1)x_2^*\\\\
x_1^*=\frac{\mu_2-\Phi(N^*)}{\mu_2-\mu_1}\\\\
x_2^*=\frac{\Phi(N^*)-\mu_1}{\mu_2-\mu_1}\\\\
\mbox{ provided that } \mu_1<\Phi(N^*)<\mu_2.
\end{array}.\eae

Notice that there exists a threshold $N_c=\sqrt{b}$, such that $\Phi(N)=\frac{r N}{b+ N^2}$ is increasing with respect to $N$ if $N<N_c$ while it is decreasing if $N>N_c$. For a given population size $N^*$ such that the condition $\mu_1<\Phi(N^*)<\mu_2$ holds, then we have the following equalities:

\bae\label{X-N}
\begin{array}{lcl}
\frac{\partial x_1^*}{\partial N^*}=\frac{-\frac{\partial \Phi(N^*)}{\partial N^*}}{\mu_2-\mu_1}\\\\
\frac{\partial x_2^*}{\partial N^*}=\frac{\frac{\partial \Phi(N^*)}{\partial N^*}}{\mu_2-\mu_1}\\\\
\mbox{ provided that } \mu_1<\Phi(N^*)<\mu_2.
\end{array}\eae
which indicates that
$$\frac{\partial x_1^*}{\partial N^*}<0 \mbox{ when } N^*<N_c; \,\,\frac{\partial x_1^*}{\partial N^*}>0 \mbox{ when } N^*>N_c$$ and
$$\frac{\partial x_2^*}{\partial N^*}>0 \mbox{ when } N^*<N_c;\,\, \frac{\partial x_2^*}{\partial N^*}<0 \mbox{ when } N^*>N_c.$$
Recall that $x_1^*$ is the allocation of the inside colony task such as brood care with the smaller mortality rate and $x_2^*$ is the allocation of the outside colony task such as foraging with the larger mortality rate. We can conclude that the colony size has important effects on the task allocation. There exists a threshold size $N_c=\sqrt{b}$: when the colony size is below this threshold, the allocation of the outside colony task such as foraging increases with the colony size. However, if the colony size is above this threshold, the allocation of the outside colony task decreases with the colony size. From biological point of view, we could consider this threshold $N_c$ is a population size that determines the stage of the colony:
\begin{enumerate}
\item If the colony size is less than this threshold, i.e., $N<N_c$, the colony is at the early ergonomic growth stage where the colony invests its resource for the colony growth. Thus, the colony requires more workers to perform the outside colony task such as foraging to support the growth need of the colony. 
\item If the colony size is larger than this threshold, i.e., $N>N_c$, the colony is at the matured stage where the colony invests its resource for the mantainence of the colony. During this stage, the colony shits workers that perform the outside colony task to perform the safer task, i.e., the inside colony task, such as brood care to take care new queens and drone. Thus, the allocation of the outside colony task is expected to be decreasing with respect the colony size.
\end{enumerate}

Our analytical results are supported by empirical findings. For example, in colonies of the common black garden ant, Lasius niger, reduced the proportion of the colony that foraged as colony size increased \cite{Gordon2010}. 
In harvest ant colonies, it is found that larger colonies increased allocation to less risky activities such as trash removal instead of more risky task such as foraging \cite{Holbrook2013b}.  When honey bee colonies were manipulated to reduce worker numbers, they immediately increased  pollen collection (foraging food for brood), and increased the number of brood being reared \cite{Seeley2009}. \\

\subsection{Effects of social interactions among task groups\\}  

Assume that social interactions (such as antennation in ants) among task groups $\{f_{ij}^T\}_{i,j=1}^m$ do not contribute to the task allocation (i.e., the model \eqref{nTL-s-nc}), then the  task allocation $X^*$ when $s=m=2$ satisfies the following equation according to \eqref{eqx-nc} and \eqref{eqx2}:
$$\begin{array}{lcl}
\left[1-x_1\right]^2&=&\frac{\mu_1\hat{D}_2}{\mu_2\hat{D}_1}x_1^2\\
\end{array}.$$

The task allocation $x_1^*$ of Model \eqref{nTL-s-nc} without social communications can be described as the following lemma:\\
\begin{lemma}[Task allocation without the contribution of social antennation interactions ]\label{l4:allocation} Let $s=m=2,\,\,\mu_1<\mu_2$, and assume that Condition \textbf{H} holds. Then the model \eqref{nTL-s-nc} has a unique task allocation $X^*=\left(x_1^*, x_2^*\right)=\left(\frac{1}{1+\sqrt{\frac{\mu_1\hat{D}_2}{\mu_2\hat{D}_1}}},\frac{\sqrt{\frac{\mu_1\hat{D}_2}{\mu_2\hat{D}_1}}}{1+\sqrt{\frac{\mu_1\hat{D}_2}{\mu_2\hat{D}_1}}}\right)$. Moreover, the following statements hold:
\begin{enumerate}
\item If $\frac{\mu_1}{\mu_2}>\frac{\hat{D}_1}{\hat{D}_2}$ holds, then the allocation of the inside task $x_1^*$ of the model \eqref{nTL-s-nc}  is smaller than its allocation of the outside task $x_2^*$. 

\item If $\frac{\mu_1}{\mu_2}<\frac{\hat{D}_1}{\hat{D}_2}$ holds, then the allocation of the inside task $x_1^*$ of the model \eqref{nTL-s-nc}  is larger than its allocation of the outside task $x_2^*$. \\
\end{enumerate}
\end{lemma}
\noindent\textbf{Notes:} The proof of Lemma \ref{l4:allocation} can be obtained through solving $x_1^*$ from the following equation:
$$\begin{array}{lcl}
\left[1-x_1\right]^2&=&\frac{\mu_1\hat{D}_2}{\mu_2\hat{D}_1}x_1^2\\\\
\end{array}.$$
Lemma \ref{l4:allocation} implies that if social interactions do not contribute to the task allocation, then the task allocation is totally determined by the ratio of the relative mortality $\frac{\mu_1}{\mu_2}$ to the relative task demand $\frac{\hat{D}_1}{\hat{D}_2}$, i.e., the inside colony task allocation is $\frac{1}{1+\sqrt{\frac{\frac{\mu_1}{\mu_2}}{\frac{\hat{D}_1}{\hat{D}_2}}}}$. This ratio could be considered as the relative cost of the inside colony task, thus  the larger the ratio, the smaller allocation to the inside colony task. \\\\

\noindent\textbf{How social interactions regulate the task allocation:}  If  social interactions do contribute to the task allocation, i.e., the full model \eqref{nTL-s}, then the allocation of the inside colony task $x_1^*$ is determined by the equation \eqref{eqx2}, i.e.,
$$\begin{array}{lcl}
\left[1-x_1\right]^2=\frac{\mu_1\hat{D}_2}{\mu_2\hat{D}_1}x_1^2+\underbrace{\frac{x_1(1-x_1)}{\mu_2}\left[\left( 1+\frac{\hat{D}_2}{\hat{D}_1}\right)x_1-1\right]}_\text{contribution from social interactions}\Rightarrow\\\\
x_2^2=\frac{\mu_1\hat{D}_2}{\mu_2\hat{D}_1}x_1^2+\underbrace{\frac{x_1(1-x_1)}{\mu_2}\left[\left( 1+\frac{\hat{D}_2}{\hat{D}_1}\right)x_1-1\right]}_\text{contribution from social interactions}
\end{array}$$
 where the contribution of social interactions to the task allocation follows the following expression:
 \bae\label{social}
SI(x_1)&=& \frac{x_1(1-x_1)}{\mu_2}\left[\left( 1+\frac{\hat{D}_2}{\hat{D}_1}\right)x_1-1\right]
 \eae which implies that
 $$SI(x_1)>0 \mbox{ when } 1>x_1>\frac{1}{1+\frac{\hat{D}_2}{\hat{D}_1}};\,\,SI(x_1)<0 \mbox{ when } x_1<\frac{1}{1+\frac{\hat{D}_2}{\hat{D}_1}}.$$
 
 Define $x^{NS}_1=\frac{1}{1+\sqrt{\frac{\mu_1\hat{D}_2}{\mu_2\hat{D}_1}}}$ as the inside colony task allocation of the task organization model \eqref{nTL-s-nc} that has no contribution from social interactions; $x^{SI}_1=\frac{1}{1+\frac{\hat{D}_2}{\hat{D}_1}}$ as a threshold determining the effects from social interactions $SI(x_1)$; and $x^R_1$ as the inside colony task allocation of the task organization model \eqref{nTL-s} that has contributions from social interactions. Then we have the following theorem regarding the effects of social interactions to the task allocation:
 
 \begin{theorem}[The contribution of social interactions to the task allocation ]\label{th4-1:allocation} Let $s=m=2,\,\,\mu_1<\mu_2$, and assume that Condition \textbf{H} holds. Then the following statements hold:
\begin{enumerate}
\item If $\frac{\mu_1}{\mu_2}>\frac{\hat{D}_2}{\hat{D}_1}$ holds, then the allocation of the inside task $x_1^{NS}$ of the model \eqref{nTL-s-nc} without social interactions is smaller than the allocation of the inside task $x_1^R$ of the model \eqref{nTL-s} with social interactions, i.e., 
$$x_1^{NS}<x_1^R<x_1^{SI}.$$

\item If $\frac{\mu_1}{\mu_2}<\frac{\hat{D}_2}{\hat{D}_1}$ holds, then the allocation of the inside task $x_1^{NS}$ of the model \eqref{nTL-s-nc} without social interactions is larger than the allocation of the inside task $x_1^R$ of the model \eqref{nTL-s} with social interactions, i.e., 
$$x_1^{SI}<x_1^R<x_1^{NS}.$$

\end{enumerate}
\end{theorem}
\noindent\textbf{Notes:} Theorem \ref{th4-1:allocation} implies that the effect of social interactions is  determined by $\frac{\frac{\mu_1}{\mu_2}}{\frac{\hat{D}_2}{\hat{D}_1}}=\frac{\mu_1\hat{D}_1}{\mu_2\hat{D}_2}$ which is the ratio of the product of the mortality and the relative demand of the inside colony task to the  product of the mortality and the relative demand of the inside colony task:
\begin{enumerate}
\item If the ratio $\frac{\mu_1\hat{D}_1}{\mu_2\hat{D}_2}$ is greater than 1, i.e. $\frac{\mu_1}{\mu_2}>\frac{\hat{D}_2}{\hat{D}_1}$, then social interactions would regulate the outside colony workers back to the colony to perform the inside colony task. Note that $\mu_1<\mu_2$, thus we have the relative demand of the outside colony task is less than the inside colony task, $\frac{\hat{D}_2}{\hat{D}_1}<1$. Therefore, social interactions among workers could rearrange more workers for the inside colony task to satisfy its demand.   
\item If the ratio $\frac{\mu_1\hat{D}_1}{\mu_2\hat{D}_2}$ is less than 1, i.e. $\frac{\mu_1}{\mu_2}<\frac{\hat{D}_2}{\hat{D}_1}$, then social interactions would regulate more workers to perform the outside colony task due to the large relative demand $\frac{\hat{D}_2}{\hat{D}_1}$ of the outside colony task. \\
\end{enumerate}

\section{The task organization model with temporal polyethism}
Temporal polyethism is ubiquitous among eusocial insect colonies \cite{Wilson1971}. It is a mechanism of task allocation where tasks in a colony are allocated among workers based on their age. In general, newly emerged workers perform less ricky tasks within the nest, such as brood care and nest maintenance, and progress to more risky tasks outside the nest, such as foraging, nest defense, and corpse removal as they age. For example, in honeybees, the youngest workers (from about 1-3 days) exclusively clean cells. From about 3-11 days, workers perform tasks related to brood care and nest maintenance. From 11-20 days, they transition to receiving and storing food from foragers, and at about 20 days workers begin to forage \cite{Seeley1982}.  Similar temporal polyethism patterns can be seen in primitive species of wasps and many species of ants where young workers feed larvae, and then transition to nest building tasks, followed by foraging \cite{Naug1998,Holldobler1990}. However, this pattern is not rigid. Workers of certain ages have strong tendencies to perform certain tasks, but may switch to other tasks through social interactions when the stimulus of other tasks are high. For instance,  removing young workers from the nest of the ant Pheidole dentata will cause foragers, especially younger foragers, to revert to tasks such as caring for brood \cite{Muscedere2009}\\

There are limited mathematical models on task organizations of social insect colonies with temporal polyethism (but see \cite{Wakano1998}). In this section, we extend  the model \eqref{TL-s} to the following model \eqref{TL-a} to incorporate temporal polyethism:
\bae\label{TL-a}
\begin{array}{lcl}
N'&=& \frac{r N^s}{b+N^s}-\sum_{j=1}^m \mu_j T_j\\\\
T_1'&=&\frac{r N^2}{b+ N^2}+\frac{T_1}{N}\sum_j\frac{\left(\frac{D_1}{\theta_1} -\frac{D_j}{\theta_j}\right)T_j}{\sum_k \frac{D_k}{\theta_k}}-(\mu_1+\beta_{12})T_1\\\\
T_i'&=&\beta_{i-1i}T_{i-1}+\frac{T_i}{N}\sum_j\frac{\left(\frac{D_i}{\theta_i} -\frac{D_j}{\theta_j}\right)}{\sum_k \frac{D_k}{\theta_k}}T_j
-\left[\mu_i+\beta_{ii+1}\right]T_i,\,\, 2\leq i\leq m-1\\\\
T_m'&=&\beta_{m-1m}T_{m-1}+\frac{T_m}{N}\sum_j\frac{\left(\frac{D_m}{\theta_m} -\frac{D_j}{\theta_j}\right)}{\sum_k \frac{D_k}{\theta_k}}T_j
-\mu_1T_m, \\\\
D_i'&=&\gamma_i N-\alpha_{i} T_i D_i.
\end{array}
\eae The modeling assumptions are similar to the model \eqref{TL-s} introduced in Section 2, except that we assume that newborn workers are born in the task status $i= 1$, e.g, brood care, with the rate $\frac{r N^2}{b+ N^2}$ while the population size of the task group $i\geq 2$ is determined by the maturation rate $\beta_{i-1i}$ from the task group $i-1$ to the task group $i$. Let $x_i=\frac{T_i}{N}$, the model \eqref{TL-a} can be rewritten as follows:
\bae\label{nTL-a}
\begin{array}{lcl}
N'&=&N\left[ \frac{r N^{s-1}}{b+N^s}-\sum_{j=1}^m \mu_j x_j\right]\\\\
x_1'&=&\frac{r N^{s-1}}{b+ N^s}+x_1\sum_j\frac{\left(\frac{D_1}{\theta_1} -\frac{D_j}{\theta_j}\right)x_j}{\sum_k \frac{D_k}{\theta_k}}-(\mu_1+\beta_{1,2})x_1-x_1 \frac{r N^{s-1}}{b+N^s}+x_1\sum_{j=1}^m \mu_j x_j\\\\
x_i'&=&\beta_{i-1,i}x_{i-1}+x_i\sum_j\frac{\left(\frac{D_i}{\theta_i} -\frac{D_j}{\theta_j}\right)}{\sum_k \frac{D_k}{\theta_k}}x_j
-(\mu_i+\beta_{i,i+1})x_i-x_i \frac{r N^{s-1}}{b+N^s}+x_i\sum_{j=1}^m \mu_j x_j,\,\, 2\leq i\leq m-1\\\\
x_m'&=&\beta_{m-1,m}x_{m-1}+x_m\sum_j\frac{\left(\frac{D_m}{\theta_m} -\frac{D_j}{\theta_j}\right)}{\sum_k \frac{D_k}{\theta_k}}x_j
-\mu_mx_m-x_m \frac{r N^{s-1}}{b+N^s}+x_m\sum_{j=1}^m \mu_j x_j, \\\\
D_i'&=&N\left[\gamma_i -\alpha_{i} x_i D_i\right]
\end{array}.
\eae
Define $\hat{D}_i=\frac{\gamma_i}{\alpha_i\theta_i}$, then we have the following theorem:
\begin{theorem}[Basic dynamical properties with age structure related DOL]\label{th5:bp-age} Assume that all parameters are strictly positive and Condition \textbf{H} holds.
Then the model \eqref{nTL-a} is positively invariant and bounded in $\Omega$. More specifically, we have 
$$N^+_m\leq \liminf_{t\rightarrow\infty} N(t)\leq\limsup_{t\rightarrow\infty} N(t)\leq N^+_1 \mbox{ and }\frac{\gamma_i}{\alpha_i}\leq \liminf_{t\rightarrow\infty} D_i(t)\leq D_M.$$
If $N(0)<N^{-}_1$, then we have $\liminf_{t\rightarrow\infty}N(t)=0.$ Moreover,  if $(N^*, X^*, D^*)$ is an equilibrium of Model \eqref{nTL-a} where $X^*=(x_1^*,...,x^*_m),\,\, D^*=(D^*_1,...,D^*_m)$, then it satisfies the following equations:
 \bae\label{eqnxd-a}
 \begin{array}{l}
 D^*_i=\frac{\gamma_i}{\alpha_i x_i^*}\mbox{ and } \frac{r (N^*)^{s-1}}{b+ (N^*)^s}=\sum_{j=1}^m \mu_j x_j^*\\\\
\left[\sum_{j=1}^m \mu_j x_j^*\right]-x_1^*\left[\mu_1+\beta_{1,2}\right]=\frac{x_1^*\left(\sum_j\hat{D}_j\right)-\hat{D}_1}{\sum_k \frac{\hat{D}_k}{x^*_k}}\\\\
\frac{x^*_{i-1}}{x^*_i}= \left[\frac{\mu_i+\beta_{i,i+1}}{\beta_{i-1i}}+\frac{\left(\sum_j\hat{D}_j\right)-\frac{\hat{D}_1}{x_i^*}}{\beta_{i-1,i}\sum_k \frac{\hat{D}_k}{x^*_k}}\right],\,\, 1<i\leq m-1\\\\
\modifyr{\frac{x^*_{m-1}}{x^*_m}= \left[\frac{\mu_m}{\beta_{m-1,m}}+\frac{\left(\sum_j\hat{D}_j\right)-\frac{\hat{D}_m}{x_m^*}}{\beta_{m-1,m}\sum_k \frac{\hat{D}_k}{x^*_k}}\right]}\mbox{ and }\sum_{i=1}^m x_i^*=1, i=1,..,m\\\\

\end{array}
 \eae
 \end{theorem}
 \noindent\textbf{Notes:} The proof of Theorem \ref{th5:bp-age} is similar to the proof of Theorem \ref{th1:bp} and Lemma \ref{l2:eq}, thus we omit the details. In the case that social interactions do not contribute to the task organization, then Model \eqref{nTL-a} is reduced to the following one:
 
 \bae\label{nTL-a-ns}
\begin{array}{lcl}
N'&=&N\left[ \frac{r N^{s-1}}{b+N^s}-\sum_{j=1}^m \mu_j x_j\right]\\\\
x_1'&=&\frac{r N^{s-1}}{b+ N^s}-(\mu_1+\beta_{12})x_1-x_1 \frac{r N^{s-1}}{b+N^s}+x_1\sum_{j=1}^m \mu_j x_j\\\\
x_i'&=&\beta_{i-1i}x_{i-1}-(\mu_i+\beta_{i,i+1})x_i-x_i \frac{r N^{s-1}}{b+N^s}+x_i\sum_{j=1}^m \mu_j x_j,\,\, 2\leq i\leq m-1\\\\
x_m'&=&\beta_{m-1m}x_{m-1}-\mu_mx_m-x_m \frac{r N^{s-1}}{b+N^s}+x_m\sum_{j=1}^m \mu_j x_j \\\\
D_i'&=&N\left[\gamma_i -\alpha_{i} x_i D_i\right]
\end{array},
\eae whose equilibrium $(N^*, X^*, D^*)$ satisfies the following equations:
 \bae\label{nTL-a-ns-eq}
  \begin{array}{l}
 D^*_i=\frac{\hat{D}_i}{x_i^*},\,\, \Phi(N^*)=\sum_{j=1}^m \mu_j x_j^*=(\mu_1+\beta_{12})x_1^* \mbox{ and }\sum_{i=1}^m x_i^*=1\\\\
 x^*_1=\frac{\Phi(N^*)}{\mu_1+\beta_{12}},\, x^*_i= x^*_1\prod_{k=2}^{i}\frac{\beta_{k-1,k}}{\mu_k+\beta_{k,k+1}},i\leq m-1,\,x^*_m= \frac{\beta_{m-1,m}x^*_1}{\mu_m}\prod_{k=2}^{m-1}\frac{\beta_{k-1,k}}{\mu_k+\beta_{k,k+1}}
 \end{array}
 \eae where $\Phi(N^*)=\frac{r (N^*)^{s-1}}{b+(N^*)^s}.$ This gives the following lemma:
 \begin{lemma}\label{l3:eq}[Equilibrium]
 If $(N^*, X^*, D^*)$ is an equilibrium of Model \eqref{nTL-a-ns} where $X^*=(x_1^*,...,x^*_m),\,\, D^*=(D^*_1,...,D^*_m)$, then it satisfies the following equations:
 \bae\label{nTL-a-ns-eq2}
 \begin{array}{l}
 x^*_1=\frac{1}{1+\frac{\beta_{m-1,m}}{\mu_m}\prod_{k=2}^{m-1}\frac{\beta_{k-1,k}}{\mu_k+\beta_{k,k+1}}+\sum_{i=2}^{m-1}\left(\prod_{k=2}^{i}\frac{\beta_{k-1,k}}{\mu_k+\beta_{k,k+1}}\right)}\\\\
\Phi(N^*)=\frac{\mu_1+\beta_{1,2}}{1+\frac{\beta_{m-1,m}}{\mu_m}\prod_{k=2}^{m-1}\frac{\beta_{k-1,k}}{\mu_k+\beta_{k,k+1}}+\sum_{i=2}^{m-1}\left(\prod_{k=2}^{i}\frac{\beta_{k-1,k}}{\mu_k+\beta_{k,k+1}}\right)}\\\\
 D^*_i=\frac{\hat{D}_i}{x_i^*},\,\,  x^*_i= x^*_1\prod_{k=2}^{i}\frac{\beta_{k-1,k}}{\mu_k+\beta_{k,k+1}},i\leq m-1,\,x^*_m= \frac{\beta_{m-1,m}x^*_1}{\mu_m}\prod_{k=2}^{m-1}\frac{\beta_{k-1,k}}{\mu_k+\beta_{k,k+1}}
\end{array}
 \eae
\end{lemma}
\noindent\textbf{Notes:} Since $\sum_i x_i^*=1$, then the equations \eqref{nTL-a-ns-eq} imply that
$$ \begin{array}{l}
\sum_{i=1}^m x_i^*=x^*_1 \left[1+\frac{\beta_{m-1,m}}{\mu_m}\prod_{k=2}^{m-1}\frac{\beta_{k-1,k}}{\mu_k+\beta_{k,k+1}}+\sum_{i=2}^{m-1}\left(\prod_{k=2}^{i}\frac{\beta_{k-1,k}}{\mu_k+\beta_{k,k+1}}\right)\right]=1\\\\
x^*_1=\frac{1}{1+\frac{\beta_{m-1,m}}{\mu_m}\prod_{k=2}^{m-1}\frac{\beta_{k-1,k}}{\mu_k+\beta_{k,k+1}}+\sum_{i=2}^{m-1}\left(\prod_{k=2}^{i}\frac{\beta_{k-1,k}}{\mu_k+\beta_{k,k+1}}\right)}
\end{array}.$$
 In the following section, we focus on the case when $s=2$ and $m=2$.\\

\subsection{Application to the allocation of the outside colony task and the inside colony task\\}
In the rest of the section, we focus on the case when $s=2$ and $m=2$, i.e., two tasks groups (the outside colony task versus the inside colony task). Without loss of generality, we let $x_1$ be the task allocation for the inside colony task  such as brood care. This gives $x_2=1-x_1$ which is the task allocation for the outside colony task  such as foraging. In general, the younger workers perform the inside colony task that has a lower mortality rate that the older workers who perform the outside colony task due to predation, whether, etc. It is natural to assume that $\mu_1<\mu_2$. \\

 For $s=2,m=2$, then according to Theorem \ref{th5:bp-age}, the task allocation for the inside colony task $x_1^*$ satisfies the following equations:
 $$\begin{array}{l}
\left[\mu_2+(\mu_1-\mu_2)x_1\right]-x_1\left[\mu_1+\beta_{12}\right]=\frac{x_1(1-x_1)\left[x_1\left(\sum_j\hat{D}_j\right)-\hat{D}_1\right]}{\hat{D}_1(1-x_1)+\hat{D}_2 x_1}\\\\
\left[\mu_2+(\mu_1-\mu_2)x_1-x_1\left(\mu_1+\beta_{12}\right)\right]\left[\hat{D}_1(1-x_1)+\hat{D}_2 x_1\right]=x_1(1-x_1)\left[x_1\left(\hat{D}_1+\hat{D}_2\right)-\hat{D}_1\right]\\\\
\left[\mu_2-(\beta_{12}+\mu_2)x_1\right]\left[1-x_1+\frac{\hat{D}_2}{\hat{D}_1} x_1\right]=x_1(1-x_1)\left[x_1\left(1+\frac{\hat{D}_2}{\hat{D}_1}\right)-1\right]\\\\\end{array}%
 $$which gives the following equation:\\
\bae\label{eq-a}
\begin{array}{l}
(1-x_1)^2=\frac{\beta_{12}\hat{D}_2}{\hat{D}_1\mu_2}x_1^2+\frac{x_1(1-x_1)}{\mu_2}\left[\underbrace{\left(1+\frac{\hat{D}_2}{\hat{D}_1}\right)x_1-1}_\text{social interaction}+\beta_{12}-\frac{\hat{D}_2\mu_2}{\hat{D}_1}\right]\\\\
 \end{array}.%
 \eae 
Recall that the task allocation $x_1^*$ of the model without age structure \eqref{nTL-s} satisfies the following equation:
$$\begin{array}{lcl}\left[1-x_1\right]^2&=&\frac{\mu_1\hat{D}_2}{\mu_2\hat{D}_1}x_1^2+\underbrace{\frac{x_1(1-x_1)}{\mu_2}\left[\left( 1+\frac{\hat{D}_2}{\hat{D}_1}\right)x_1-1\right]}_\text{contribution from social interactions}.
\end{array}$$ This implies follows the task allocation of Model \eqref{nTL-a} does not depend on the mortality of the inside colony task $\mu_1$, instead, the maturation rate $\beta_{1,2}$ from the inside colony task group to the outside colony task group, the  mortality of the outside colony task $\mu_2$, and the relative demand of the outside colony task $\frac{\hat{D}_2}{\hat{D}_1}$ have important effects.  Now we claim the following theorem:\\
 
 \begin{theorem}[Task allocation]\label{th5:allocation} Let $s=m=2,\,\,\mu_1<\mu_2$, and  assume that Condition \textbf{H} holds. Then the model \eqref{nTL-a} has two interior equilibria $(N^*_\pm,X^*,D^*)=\left(\frac{r\pm\sqrt{r^2-4b\mu^2}}{2\mu}, x_1^*,1-x_1^*, \frac{\gamma_1}{x_1^*\alpha_1} , \frac{\gamma_2}{(1-x_1^*)\alpha_2}\right)$ where $\mu=\mu_1x_1^*+\mu_2(1-x_1^*)$, and $x_1^*\in(0,1)$ which is the positive root of \eqref{eq-a}. Moreover, the following statements hold:
\begin{enumerate}
\item The equilibrium $(N^*_-,X^*,D^*)$ is unstable, the total population $N^*_\pm$ are in the following intervals:
$$N^*_+\in\left(\frac{r+\sqrt{r^2-4b\mu^2_2}}{2\mu_2},\frac{r+\sqrt{r^2-4b\mu^2_1}}{2\mu_1}\right)\mbox{ and } N^*_-\in\left(\frac{r-\sqrt{r^2-4b\mu^2_1}}{2\mu_1},\frac{r-\sqrt{r^2-4b\mu^2_2}}{2\mu_2}\right).$$ 
In addition, $N^*_+$ is  an increasing function of $x_1^*$, while it is a decreasing function of $x_2^*, \beta_{1,2}, \mu_1$, respectively. 
\item If $\beta_{1,2}<1+\frac{\mu_2\hat{D}_2}{\hat{D}_1}$, then the allocation of the inside colony task $x_1^*$ is greater than the critical value $\hat{x}_1$ where $\hat{x}_1=\frac{a_1-\sqrt{a_1^2-4a_0}}{2}\in (0,1)$ with
$$a_1=1+\frac{\frac{\hat{D}_2(\mu_2+\beta_{1,2})}{\hat{D}_1}+1-\beta_{1,2}}{1+\frac{\hat{D}_2}{\hat{D}_1}},\,a_0=\frac{1+\frac{\mu_2\hat{D}_2}{\hat{D}_1}-\beta_{1,2}}{1+\frac{\hat{D}_2}{\hat{D}_1}}.$$

\item The allocation of the inside colony task $x_1^*$  is decreasing with respect to $\beta_{1,2}$. If $\beta_{1,2}>1+\frac{\mu_2\hat{D}_2}{\hat{D}_1}$, then $x_1^*$ is increasing with respect to $\mu_2$.\\
\end{enumerate}
\end{theorem}
 \noindent\textbf{Notes:}  The maturation rate $\beta_{1,2}$ from the inside colony task group to the outside colony task group could be considered as the additional ``death" rate of the  inside colony task group and the birth rate of the  outside colony task group.  Theorem \ref{th5:allocation}  implies that the relative small value of $\beta_{1,2}$ (e.g., $\beta_{1,2}<1+\frac{\mu_2\hat{D}_2}{\hat{D}_1}$) can guarantee the inside task allocation $x_1^*$ above some threshold $\hat{x}_1$, however, if $\beta_{1,2}$  is too large, then the inside task allocation $x_1^*$  would be very small since it is decreasing with respect to $\beta_{1,2}$. These theoretical results suggest that the timing of worker maturation may be important in shaping various aspects of task organizations. For example, the empirical study of honey bees by Giray \emph{et al} \cite{Giray2000} support that faster individual behavioral development  may result in a larger force of foragers and a smaller force of nurse. In addition, Theorem \ref{th5:allocation} indicates that when the maturation rate $\beta_{1,2}$ is large, then increasing the average mortality of the outside colony task can increase the inside colony task allocation. Its biological implication is that the colony will regulate the outside colony workers back to the colony to perform less risky work when the cost of the outside colony task is too high.\\

 \noindent\textbf{Effects of the colony size $N$:} Similar to our discussion provided in Section 3.1 and 3.2, \textbf{the matured colony} (i.e., the size is above a threshold) distributes more allocation to the inside colony task distribution $x_1^*$ (e.g., invests more in less risky task) as its size increases, while \textbf{the early stage colony} (i.e., the size is below a threshold) allocate more force to the outside colony task distribution $x_2^*$ for growth.
 In addition, \textbf{the matured colony} is decreasing with respect to the mortality of the  inside colony task $\mu_1$ and the maturation rate $\beta_{1,2}$.\\



\subsection{Effects of social interactions among task groups\\}  

Assume that social interactions (such as antennation in ants) among task groups $\{f_{ij}^T\}_{i,j=1}^m$ do not contribute to the task allocation (i.e., the model \eqref{nTL-a-ns}), then the  task allocation $X^*=(x^*_1,1-x^*_2)=\left(\frac{\mu_2}{\beta_{1,2}+\mu_2},\frac{\beta_{1,2}}{\beta_{1,2}+\mu_2}\right)$ which satisfies the following equation according to Lemma \ref{l3:eq}
$$\begin{array}{l}
  (1-x_1)^2=\frac{\beta_{12}\hat{D}_2}{\hat{D}_1\mu_2}x_1^2+\frac{x_1(1-x_1)}{\mu_2}\left[\beta_{12}-\frac{\hat{D}_2\mu_2}{\hat{D}_1}\right].
   \end{array}$$

If  social interactions do contribute to the task allocation, i.e., the full model \eqref{nTL-a}, then the allocation of the inside colony task $x_1^*$ is determined by the equation \eqref{eq-a}, i.e.,
$$\begin{array}{l}
(1-x_1)^2=\frac{\beta_{12}\hat{D}_2}{\hat{D}_1\mu_2}x_1^2+\frac{x_1(1-x_1)}{\mu_2}\left[\underbrace{\left(1+\frac{\hat{D}_2}{\hat{D}_1}\right)x_1-1}_\text{social interaction}+\beta_{12}-\frac{\hat{D}_2\mu_2}{\hat{D}_1}\right]\\\\
 \end{array}$$
 where the contribution of social interactions to the task allocation follows the equation \eqref{social}.\\
 
Define $x^{NSA}_1=\frac{\mu_2}{\beta_{1,2}\mu_2}$ as the inside colony task allocation of the task organization model \eqref{nTL-a-ns} that has no contribution from social interactions; $x^{SI}_1=\frac{1}{1+\frac{\hat{D}_2}{\hat{D}_1}}$ as a threshold determining the effects from social interactions $SI(x_1)$  \eqref{social}; and $x^{RA}_1$ as the inside colony task allocation of the task organization model \eqref{nTL-a} that has contributions from social interactions. Then we have the following corollary regarding the effects of social interactions to the task allocation by applying the proof of Theorem \ref{th4-1:allocation}:\\
 
\begin{corollary}[The contribution of social interactions]\label{c6:allocation} Let $s=m=2,\,\,\mu_1<\mu_2$, and assume that Condition \textbf{H} holds. Then the model \eqref{nTL-a-ns} has a unique task allocation $X^*=\left(\frac{\mu_2}{\beta_{1,2}+\mu_2},\frac{\beta_{1,2}}{\beta_{1,2}+\mu_2}\right)$. Moreover, the following statements hold:
\begin{enumerate}
\item For Model \eqref{nTL-a-ns}, the ratio of the inside colony task allocation to the outside colony task allocation is $\frac{\mu_2}{\beta_{1,2}}.$

\item If $\frac{\beta_{1,2}}{\mu_2}<\frac{\hat{D}_2}{\hat{D}_1}$ holds, then the allocation of the inside task $x_1^{NSA}$ of the model \eqref{nTL-a-ns} without social interactions is smaller than the allocation of the inside task $x_1^R$ of the model \eqref{nTL-a} with social interactions, i.e., 
$$x_1^{NSA}<x_1^{RA}<x_1^{SI}.$$

\item If $\frac{\beta_{1,2}}{\mu_2}>\frac{\hat{D}_2}{\hat{D}_1}$ holds, then the allocation of the inside task $x_1^{NS}$ of the model \eqref{nTL-a-ns} without social interactions is larger than the allocation of the inside task $x_1^R$ of the model \eqref{nTL-a} with social interactions, i.e., 
$$x_1^{SI}<x_1^{RA}<x_1^{NSA}.$$
\end{enumerate}
\end{corollary}

 \noindent\textbf{Effects of social interactions among task groups:} Corollary \ref{c6:allocation} implies that, in the absence of the contribution of social interactions to task allocation,  the task allocation is completely determined by the mortality of the outside colony task $\mu_2$ and the maturation rate (or the birth rate of the outside colony task group) $\beta_{1,2}$ where the mortality of the inside colony task $\mu_1$ and the task demand have no effects. \\
 
 We can denote $\frac{\beta_{1,2}}{\mu_2}$ as the relative population of the outside colony task group which the ratio of the birth rate of the outside colony task group to its mortality; and $\frac{\hat{D}_2}{\hat{D}_1}$ as the relative task demand of the outside colony task. Corollary \ref{c6:allocation} indicates that if the relative population is less than that the relative demand of the outside colony task group, then social interactions will regulate outside colony workers back to colony to perform the inside colony task to satisfy the demand. However, if the relative population is more than that the relative demand of the outside colony task group, then social interactions will regulate more workers to perform outside colony tasks to optimize its investment.\\  
 
\section{Discussion}
Evidence of the ecological success of social insects can be found almost everywhere  \cite{Wilson1985b}. The organizations of insect societies, such as, division of labor, task allocation, collective regulation, mass action responses, have been considered as main reasons for the success \cite{Page1990}. Colony-level patterns including the sizes of task groups, the patterns of overlap among task groups \cite{Wilson1976}, and short-term task allocation \cite{Gordon1996}, emerge from the decisions and actions of individual workers.  Increasing evidence suggests that although genetic, physiological and other aspects must be taken into account \cite{Keller2009,ODonnell1996, Page2002}, mechanistic explanations should be studied together \cite{Burd2008, Franks2009, Sumpter2010}. In this article, we develop a framework of mathematical models to explore the crucial feedback mechanisms linking both structure and dynamics of task organizations in a dynamical environment, and investigate the potential underlying processes of task organization as colony size increases.\\

Task allocation is assessed by the distribution of number of workers across tasks \cite{Holbrook2013a, Holbrook2013b, Wilson1980a}. Social insect colonies can rapidly change its task allocation between tasks in response to changes in task need and/or environment stimuli \cite{Fewell2003, ODonnell2007, Gordon2010, Tschinkel1999}, which has been driven in part by the colony's social communications, and the high connectivity of workers across task groups \cite{Jeanson2007, Wilson1980b}. Depending on the species the tasks typically include collecting food (forager), feeding and caring for the offspring (broodcare), and defending the nest against intruders or parasites (soldier). There is an extensive body of empirical work in biology studying the phenomenon of division of labor in social insect colonies (e.g., \cite{Camazine2003, Seeley2009, Dornhaus2008a,Dornhaus2008b, Pinter-Wollman2012}), and the related individual behaviors that could generate the collective division of labor (e.g., \cite{Beekman2001, Sumpter2003, Myerscough2004}). These empirical work suggested that workers might select and potentially switch tasks based on different features, including their age \cite{Robinson1992, Seid2006}, social communications with other ants \cite{Robinson1992, Sendova-Franks1995, Gordon2003}, or the internal response thresholds and the task demands \cite{Bonabeau1996, Ravary2007}. Our proposed models include these important features. \\

More precisely, our first general model incorporates three features: (a) the average internal response threshold for each task (the internal factor); (b) social network communications that could lead to task switching (the environmental factor); and (c) dynamical changes of task demands (the external factor). Workers in many social insect species exhibit \emph{age polyethism}, in which worker task preferences change with age. It is well-known that \emph{age polyethism} could constrain the ability of individuals to switch immediately between tasks. For example, P. californicus workers have a general age transition from in-nest tasks, such as brood care, to external tasks, including foraging and waste management \cite{Smith2006, Ingram2013, Clark2014}. In order to understand how \emph{age polyethism} affect task organization such as division of labor and task allocation in addition to the three features mentioned earlier, we develop an additional general model with \emph{age polyethism} that is modified from the first model. The proposed models are biologically well-defined (see Theorem \ref{th1:bp} and Theorem \ref{th5:bp-age}). We apply our general modeling framework to the cases of the inside colony task versus the outside colony task. Our analytical study of the models  provide important insights on the effects of colony size, social communication, and age related task preferences on task allocation and division of labor. \\


When the population growth dynamics of a colony occurs at the comparable time scale of the task allocation, our theoretical results (Theorem \ref{th2:allocation}) show that the relative demand and the mortality of tasks would affect the task allocation in the following way: (1) The inside colony task allocation is above a threshold that is determined by its mortality and the relative demand. (2) When the colony has higher demand of the outside task such as foraging, the colony would distribute more workers to the outside colony task despite that this task has a higher mortality. (3) To optimize the investment of the colony, the colony would decrease the distribution to a colony task when its mortality increases. For social insect colonies with age polyethism, our results (Theorem \ref{th5:allocation}) indicate that the maturation time of workers play an important role in shaping various aspects of task organizations. If the maturation rate is large, then increasing the average mortality of the outside colony task can increase the inside colony task allocation. Its biological implication is that the colony will regulate the outside colony workers back to the colony to perform less risky work when the cost of the outside colony task is too high.\\

When the time scale of the population growth dynamics is much slower than the task allocation and the related demand (i.e., the colony size can be considered as constant when study the dynamics of task allocation), the colony size has a huge impact on the task allocations. Our analysis implies that: When the colony is at the early ergonomic growth stage (i.e., its size is below a threshold),  the colony invests its resource for the colony growth. As a consequence, the colony requires more workers to perform the outside colony task such as foraging to support the growth need of the colony. When the colony is at the matured stage (i.e., its size is above a threshold), the colony shifts more workers to perform the safer task (e.g., the inside colony task). As a consequence, the allocation of the outside colony task decreases with respect the colony size. For colonies with age polyethism, the size of the matured colony decreases when the mortality of the inside colony task (or the maturation rate of the outside colony task) increases. \\

Our theoretical findings are supported by empirical studies that have showed that the social context, in particular colony size, influences the ergonomics of insect societies. For instance, colony size shapes the exploratory and foraging responses in ants, and an increase in worker number triggers the formation of more efficient foraging networks \cite{Beekman2001}. In wasps, the delay experienced by workers during transfer of materials for nest construction decreases with group size due to the reduction of stochastic fluctuations \cite{Anderson1999}. Increased colony size can also affect the rate of interactions among workers and thus contribute significantly to the efficient allocation of individuals to different tasks \cite{Pacala1996, Naug1998}. For the harvester ant (Pogonomyrmex badius), 
in spring, proportion foraging increased due to an increase in forager number and reduction in colony size, and in late summer, it decreased as
colony size increased through new worker birth and a loss of about 3 \% of foragers per day \cite{Kwapich2013}. Moreover, immature colonies (less than 800 workers) reached a higher maximum proportion foraging than large, mature colonies each year \cite{Kwapich2013}.\\ %

Social communication in social insect colonies has profound impacts on the organization of task allocation and division of labor \cite{Charbonneau2013, Fewell2003, Gordon2010}. Empirical work show that direct physical contacts (e.g., antennations) can generate changes in the number of workers performing a task \cite{Gordon1999a,Gordon1999b,Pacala1996, Pinter-Wollman2011}, and almost all work activities, from task switching to recruitment, to assessment of task needs, require workers to interact with others. Our analysis (Lemma \ref{l4:allocation}) reveals that without the contribution from social interactions, the task allocation is totally determined by the ratio of the relative mortality  to the relative task demand. This ratio could be considered as the relative cost of the task, thus the larger the ratio, the smaller allocation to the task. In the presence of the social interaction, our results (Theorem \ref{th4-1:allocation}) imply that the effect of social interactions is  determined by the ratio of the product of the mortality and the relative demand of the inside colony task to the  product of the mortality and the relative demand of the inside colony task. If the ratio is greater than 1,  then social interactions would regulate the outside colony workers back to the colony to perform the inside colony task. 
However, if the ratio is less than 1, then social interactions would regulate more workers to perform the outside colony task due to the large relative demand  of the outside colony task. \\

For the model with temporal polyethism, our work (Corollary \ref{c6:allocation}) implies that, in the absence of the contribution of social interactions,  the task allocation is completely determined by the mortality of the outside colony task and the maturation rate (or the birth rate of the outside colony task group)  where the mortality of the inside colony task  and the task demand have no effects. With the contribution of social interaction, our analysis indicates that if the relative population is less than that the relative demand of the outside colony task group, then social interactions will regulate outside colony workers back to colony to perform the inside colony task to satisfy the demand. However, if the relative population is more than that the relative demand of the outside colony task group, then social interactions will regulate more workers to perform outside colony tasks to optimize its investment.\\

It has been shown that the efficient organization of work in  eusocial insects has attributed greatly in their outstanding ecological success \cite{Oster1978}. A prominent pattern of colony organization is division of labor, expressed not only between reproductive and worker castes, but also among workers that specialize on different tasks \cite{Wilson1971, Holldobler1990, Wilson2009}. Non-reproductive division of labor, or polyethism, is thought to enhance colony performance and ultimately fitness \cite{Oster1978, ChittkaMuller2009} however see \cite{Dornhaus2008a}. Colonies must balance the putative benefits of task specialization with the flexibility to reallocate workers in response to changes in task demand, which depends on variable internal and external factors \cite{Calabi1989,  Gordon1996, Seeley2009, Holbrook2013a}. The study present in this paper elucidates how individual task decisions are integrated with a fundamental colony attribute (colony size) and a simple behavior rule (local social interaction) by using the framework of our proposed models.\\

It would be interesting to include inactive workers (i.e., workers don't appear to do anything at all) as an additional component in our models. The presence of inactive workers in social insect colonies is an another phenomenon in social insect task allocation even in the field \cite{Jandt2009,Charbonneau2015}, which may be driven in part by selfish interests, selfishness can at most explain a small fraction of observed inactivity \cite{Jandt2011}. The recent work shows that nor do all inactive workers appear to be reserves for defense \cite{Jandt2012} or for the case of worker loss \cite{Pinter-Wollman2012}. The work of Pinter-Wollman \emph{et al.} \cite{Pinter-Wollman2012} suggests that inactivity may be a side-effect of an imperfect mechanism for generating individual variation for the purpose of effective task allocation, which is a mathematically difficult problem \cite{Cornejo2014}. We have an on-going project to investigate potential mechanisms producing inactive workers in a separate paper. \\

\section{Proofs}

\subsection*{Proof of Lemma \ref{l1:bp}}
\begin{proof}
Since $f(N,\mu)=\mu(b+N^s)-rN^{s-1}$, then simple calculations give follows:
$$f(0,\mu)=b\mu>0,\mbox{ and }\frac{\partial f(N,\mu)}{\partial N}=N^{s-2}\left[s\mu N- r(s-1)\right]$$which implies that
$f(N,\mu)$ has a unique positive critical point $N_c(\mu)=\frac{r(s-1)}{\mu s}$ and
$$\frac{\partial f}{\partial N}<0 \mbox{ if } N<N_c(\mu);\,\, \frac{\partial f}{\partial N}>0 \mbox{ if } N>N_c(\mu).$$
Therefore, if $f(N_c(\mu),\mu)<0$, the it has two positive roots $N^\pm$ such that $f(N^\pm,\mu)=0$ with $N^-<N_c(\mu)<N^+$.\\

On the other hand, we have 
$$f(N_c(\mu),\mu)=b\mu+N_c(\mu)^{s-1}(\mu N_c(\mu)-r) \mbox{ and } \frac{\partial f(N_c(\mu),\mu)}{\partial \mu}=b+N_c(\mu)^s>0$$which indicates that 
$$f(N_c(\mu_1),\mu_1)\leq f(N_c(\mu_m),\mu_m) \mbox{ for } \mu_1\leq \mu_m.$$ Therefore, if $f(N_c(\mu_m),\mu_m)<0$ then we have $f(N_c(\mu_1),\mu_1)<0$. Therefore, if $f(N_c(\mu_m),\mu_m)<0$, then there exists $N^{\pm}_1>0$ and $N^{\pm}_m>0$ such that
$$ f(N^{\pm}_1,\mu_1)=f(N^{\pm}_m,\mu_m)=0 \mbox{ and } N^{-}_1<N_c(\mu_1)<N^{+}_1, N^{-}_m<N_c(\mu_m)< N^{+}_m.$$ Since $f(N,\mu)$ is an increasing function with respect to $\mu$, thus, we can conclude that  $N^{-}_1<N^{-}_m<N^{+}_m\leq N^{+}_1.$ Therefore, the statement holds.\\

\end{proof}

\subsection*{Proof of Theorem \ref{th1:bp}}
\begin{proof}For any initial condition taken in $\Omega$, we have follows:

$$\frac{dN}{dt} \Big\vert_{N=0}=0,\,\,\frac{dx_i}{dt} \Big\vert_{x_i=0}=\frac{r N^{s-1}}{b+ N^s}\frac{\frac{D_i}{\theta_i}}{\sum_k \frac{D_k}{\theta_k}}\geq0,\,\,\frac{dD_i}{dt} \Big\vert_{D_i=0}=\gamma_i>0, i=1,..,m$$ which implies that the model \eqref{nTL-s} is positively invariant in $\Omega$ according to Theorem A.4 (p.423) in \cite{thieme2003mathematics}.\\

 Since $\sum_{i=1}^m x_i(0)=1$ and $\sum_{i=1}^m x_i'=0$, we can conclude that 
$$\sum_{i=1}^m x_i(t)=1 \mbox{ and } 0\leq x_i(t)\leq 1 \mbox{ for all } t\geq 0.$$
This implies that
$$\begin{array}{lcl}
N'&=&N\left[\frac{r N^{s-1}}{b+N^s}-\sum_{j=1}^m \mu_j x_j \right]\leq N\left[\frac{r N^{s-1}}{b+N^s}- \mu_1 \right]\\\\
N'&=&N\left[\frac{r N^{s-1}}{b+N^s}-\sum_{j=1}^m \mu_j x_j \right]\geq N\left[\frac{r N^{s-1}}{b+N^s}- \mu_m \right]\\\\
D_i'&=&N\left[\gamma_i -\alpha_{i} x_i D_i\right]\geq N\left[\gamma_i -\alpha_{i}  D_i\right]
\end{array}.$$ 
Assume that all parameters are strictly positive and Condition \textbf{H} holds. Then according to Lemma \ref{l1:bp}, there exists $N^{\pm}_1$ and $N^{\pm}_m$ such that
$$f(N^{\pm}_1,\mu_1)=f(N^{\pm}_m,\mu_m)=0 \mbox{ and } N^{-}_1<N_c(\mu_1)<N^{+}_1, N^{-}_m<N_c(\mu_m)< N^{+}_m, N^{-}_1<N^{-}_m<N^{+}_m\leq N^{+}_1.$$
This implies that if $N(0)<N^{-}_1$, then we have
$$N'\leq N\left[\frac{r N^{s-1}}{b+N^s}- \mu_m \right]<0\Rightarrow \liminf_{t\rightarrow\infty}N(t)=0.$$
If $N(0)>N^{-}_m$, then we have follows by applying inequalities:
$$0<N\left[\frac{r N^{s-1}}{b+N^s}- \mu_1 \right]\leq N'\leq N\left[\frac{r N^{s-1}}{b+N^s}- \mu_1 \right]\Rightarrow N^+_m\leq \liminf_{t\rightarrow\infty}N(t)\leq \limsup_{t\rightarrow\infty}N(t)<N^+_1.$$
If Condition \textbf{H} holds, then the inequality $D_i'\geq N^+_m\left[\gamma_i -\alpha_{i}  D_i\right]$ holds which implies that  $\liminf_{t\rightarrow\infty}D_i(t)\geq \frac{\gamma_i}{\alpha_i}$. On the other hand, we have $D_i'\leq N^+_1\gamma_i $ which gives $D_i(t)\leq \gamma_it +D_i(0).$ Therefore, we have:
$$\frac{\frac{D_i}{\theta_i}}{\sum_k \frac{D_k}{\theta_k}}\geq \frac{\frac{\gamma_i}{\alpha_i\theta_i}}{\sum_k \frac{\gamma_kt +D_k(0)}{\theta_k}}=\frac{\frac{\gamma_i}{\alpha_i\theta_i}}{\gamma t+a}.$$
Let $M=\min_{N^{+}_m\leq N\leq N^{+}_1}\{\frac{r N^{s-1}}{b+N^s}\}$, then we have
$$
\begin{array}{lcl}
\frac{dx_i}{dt}&=&\frac{r N^{s-1}}{b+N^s} \left[\frac{\frac{D_i}{\theta_i}}{\sum_k \frac{D_k}{\theta_k}}-x_i\right]+\frac{x_i\left[ \frac{D_i}{\theta_i}-\sum_j \frac{D_j}{\theta_j} x_j\right] }{\sum_k \frac{D_k}{\theta_k}} -\mu_ix_i+x_i\sum_{j=1}^m \mu_j  x_j\\\\
&\geq& M\left[\frac{\frac{\gamma_i}{\alpha_i\theta_i}}{\gamma t+a}-x_i\right]+x_i\left[\frac{\frac{\gamma_i}{\alpha_i\theta_i}}{\gamma t+a}-1-\mu_i+\mu_1\right]
\end{array}$$which implies that $x_i$ is persistent, i.e., there exists $\epsilon_i>0$ such that:
$$\liminf_{t\rightarrow\infty}x_i(t)\geq \epsilon \mbox{ for all } i=1,...,m.$$
This indicates that $D_i'\leq N^+_1\left[\gamma_i -\alpha_{i} \epsilon_i D_i\right]$ holds. Therefore, we have $\limsup_{t\rightarrow\infty}D_i(t)\leq \frac{\gamma_i}{\alpha_i\epsilon}<D_M$ for all $i=1..m$.
\end{proof}

\subsection*{Proof of Theorem \ref{th2:allocation}}
\begin{proof}
Let $f_1(x_1)=\left[1-x_1\right]^2$ where $f_1(x_1)\geq0$ is a decreasing function in $x_1$ when $x_1\in [0,1]$ and $f_1(0)=1, f_1(1)=0$. And let $$f_2(x_1)=\frac{\mu_1\hat{D}_2}{\mu_2\hat{D}_1}x_1^2+\frac{x_1(1-x_1)}{\mu_2}\left[\left( 1+\frac{\hat{D}_2}{\hat{D}_1}\right)x_1-1\right].$$
We denote $f(x_1)$ as follows:
$$f(x_1)=f_1(x_1)-f_2(x_1)=c_3 x_1^3+c_2 x_2^2+c_1x_1+c_0$$where
$$\begin{array}{lcl}
c_3=\frac{1+\frac{\hat{D}_2}{\hat{D}_1}}{\mu_2}, && c_0=1\\\\
c_2=\frac{(\mu_2-2)-\frac{\hat{D}_2}{\hat{D}_1}(1+\mu_1)}{\mu_2},&&c_1=\frac{1-2\mu_2}{\mu_2}
\end{array}$$ with $f(0)=c_0=1>0, f(1)=-\frac{\mu_1\hat{D}_2}{\mu_2\hat{D}_1}<0.$ \\
 
Notice that the function $f(x_1)$ is a polynomial with degree 3,  the coefficient of the degree 3 is  $c_3=\frac{1+\frac{\hat{D}_2}{\hat{D}_1}}{\mu_2}>0$, $f(0)=c_0=1>0, f(1)=-\frac{\mu_1\hat{D}_2}{\mu_2\hat{D}_1}<0.$ This implies that has one negative root and two positive roots where one is greater than 1 and the other is between $(0,1)$. Therefore,  $f(x_1)$ has a unique positive root $x_1^*\in (0,1)$. This implies that if the equilibrium $(N^*,X^*,D^*)$ exists, then  $(X^*,D^*)=\left(x_1^*,1-x_1^*, \frac{\gamma_1}{x_1^*\alpha_1} , \frac{\gamma_2}{(1-x_1^*)\alpha_2}\right)$; and $N^*_\pm=\frac{r\pm\sqrt{r^2-4b\mu^2}}{2\mu}$ with $\mu=\mu_1x_1^*+\mu_2(1-x_1^*)$.  Since Condition \textbf{H} holds and $\mu_1<\mu_2$, then according to Lemma \ref{l1:bp} and Theorem \ref{th1:bp}, we can conclude that $N^*_+\in \left(\frac{r+\sqrt{r^2-4b\mu^2_2}}{2\mu_2},\frac{r+\sqrt{r^2-4b\mu^2_1}}{2\mu_1}\right)$ and $N^*_-\in\left(\frac{r-\sqrt{r^2-4b\mu^2_1}}{2\mu_1},\frac{r-\sqrt{r^2-4b\mu^2_2}}{2\mu_2}\right)$. Moreover, we have
$$\frac{\partial N^*_+}{\partial \mu}=-\frac{r(r+ \sqrt{r^2-4b\mu^2})}{2\mu^2 \sqrt{r^2-4b\mu^2}}<0,\,\, \frac{\partial \mu}{\partial x^*_1}= \mu_1-\mu_2<0,\,\, \frac{\partial \mu}{\partial \mu_1}=x_1^*>0,\mbox{ and } \frac{\partial \mu}{\partial \mu_2}=1-x_1^*>0$$which implies that $N^*_+$ is an increasing function of $x_1^*$, and  it is a decreasing function of $x_2^*$, respectively. \\

According to Lemma \ref{l1:bp} and Theorem \ref{th1:bp}, we know that the total population $N$ of the model \eqref{nTL-s} converges to 0 whenever $N<N^*_-$, this implies that the equilibrium $(N^*_-,X^*,D^*)$ is unstable.\\

Let $rd=\frac{\hat{D}_2}{\hat{D}_1}$. Recall that $$f_2(x_1)=\frac{\mu_1\hat{D}_2}{\mu_2\hat{D}_1}x_1^2+\frac{x_1(1-x_1)}{\mu_2}\left[\left( 1+\frac{\hat{D}_2}{\hat{D}_1}\right)x_1-1\right]=-\frac{(1+rd)x_1}{\mu_2}\left[ x_1^2-b_1x_1+b_0\right]$$ where $b_1=1+\frac{1+rd\mu_1}{1+rd}$ and $b_0=\frac{1}{1+rd}$.  Since $b_1^2>4b_0>0$, we can conclude that $f_2(x_1)$ has zero and two positive number $\hat{x}_i,i=1,2$ where $\hat{x}_1+\hat{x}_2=b_1>1$ and $\hat{x}_1\hat{x}_2=b_0<1$. This implies that $f_2(x_1)$ has a unique positive root $\hat{x}_1=\frac{b_1-\sqrt{b_1^2-4b_0}}{2}=\frac{rd(1+\mu_1)+2-\sqrt{rd^2(1+\mu_1)^2+4rd\mu_1}}{2(1+rd)}\in (0,1)$ where
$$f_2(x_1)<0 \mbox{ if } 0<x_1<\hat{x}_1 \mbox{ and } f_2(x_1)>0 \mbox{ if } \hat{x}_1 <x_1<1.$$
Since $f_1$ is strictly positive and decreasing in $(0,1)$ and $f=f_1-f_2$ has a unique positive root $x_1^*\in (0.1)$, therefore, we can conclude that $x_1^*>\hat{x}_1$.\\

In addition,  the function $f_2(x_1)$ is an increasing function with respect to $rd=\frac{\hat{D}_2}{\hat{D}_1}$, $\mu_1$, and is decreasing with respect to $\mu_2$. This implies that increasing the value of $rd$ or $\mu_1$, the intercept of $f_2(x_1)=f_1(x_1)$ is decreasing, i.e., the unique positive solution $x^*_1$ of $f_2(x_1)=f_1(x_1)$ is decreasing with respect to $rd$, and $\mu_1$, respectively. Similarly, we can conclude that the unique positive solution $x^*_1$ of $f_2(x_1)=f_1(x_1)$ is increasing with respect to $\mu_2$. This also implies that
$$\frac{d N^*_+}{d \mu_1}=\frac{\partial N^*_+}{\partial \mu}\frac{\partial \mu}{\partial \mu_1}+\frac{\partial N^*_+}{\partial \mu}\frac{\partial \mu}{\partial x^*_1}\frac{\partial x_1^*}{\partial \mu_1}<0$$ and
$$\frac{\partial N^*_+}{\partial rd}=\frac{\partial N^*_+}{\partial \mu}\frac{\partial \mu}{\partial x^*_1}\frac{\partial x_1^*}{\partial rd}<0.$$
Therefore, the total population $N^*_+$ is a decreasing function of $\mu_1$ and $rd=\frac{\hat{D}_2}{\hat{D}_1}$, respectively. Hence, the statements hold.\\
\end{proof}





\subsection*{Proof of Theorem \ref{th4-1:allocation}\\}
\begin{proof}

Let $x^{NS}_1=\frac{1}{1+\sqrt{\frac{\mu_1\hat{D}_2}{\mu_2\hat{D}_1}}}$ be the inside colony task allocation of the task organization model \eqref{nTL-s-nc} that has no contribution from social interactions, then we have
$$\left[1-x^{NS}_1\right]^2=\frac{\mu_1\hat{D}_2}{\mu_2\hat{D}_1}(x^{NS}_1)^2.$$
Let $x^{SI}_1=\frac{1}{1+\frac{\hat{D}_2}{\hat{D}_1}}$ be a positive root of the contribution function $SI(x_1)$ of social interactions to the task allocation, i.e., $$SI(x_1)= \frac{x_1(1-x_1)}{\mu_2}\left[\left( 1+\frac{\hat{D}_2}{\hat{D}_1}\right)x_1-1\right].$$
 Then we have 
 $$SI(x_1)>0 \mbox{ when } 1>x_1>x^{SI}_1;\,\,SI(x_1)<0 \mbox{ when } x_1<x^{SI}_1.$$
Based on the proof of Theorem  \ref{th2:allocation}, we have $f_1(x_1)=\left[1-x_1\right]^2$ where $f_1(x_1)\geq0$ is a decreasing function in $x_1$ when $x_1\in [0,1]$ and $f_1(0)=1, f_1(1)=0$; and $f_2(x_1)=\frac{\mu_1\hat{D}_2}{\mu_2\hat{D}_1}x_1^2+SI(x_1)$ with $f_2(0)=0;\,f_2(1)=\frac{\mu_1\hat{D}_2}{\mu_2\hat{D}_1}$.  Let  $x^R_1$ be the inside colony task allocation of the task organization model \eqref{nTL-s} that has contributions from social interactions, then it is a unique positive root in $(0,1)$ of the following equation
$$f_1(x_1)=f_2(x_1)=\frac{\mu_1\hat{D}_2}{\mu_2\hat{D}_1}x_1^2+SI(x_1).$$
If  $\frac{\mu_1}{\mu_2}>\frac{\hat{D}_2}{\hat{D}_1}$, then we can conclude that 
$$1+\sqrt{\frac{\mu_1\hat{D}_2}{\mu_2\hat{D}_1}}>1+\frac{\hat{D}_2}{\hat{D}_1}\Leftrightarrow x_1^{NS}=\frac{1}{1+\sqrt{\frac{\mu_1\hat{D}_2}{\mu_2\hat{D}_1}}}<x_1^{SI}=\frac{1}{1+\frac{\hat{D}_2}{\hat{D}_1}}.$$
This also implies that 
$$f_2(x_1)=\frac{\mu_1\hat{D}_2}{\mu_2\hat{D}_1}x_1^2+SI(x_1)<\frac{\mu_1\hat{D}_2}{\mu_2\hat{D}_1}x_1^2 \mbox{ when } x_1<x^{SI}_1$$ and
$$f_2(x_1)=\frac{\mu_1\hat{D}_2}{\mu_2\hat{D}_1}x_1^2+SI(x_1)>\frac{\mu_1\hat{D}_2}{\mu_2\hat{D}_1}x_1^2 \mbox{ when } 1>x_1>x^{SI}_1>x^{NS}_1.$$
Notice that $f_1(x_1)$ is decreasing in $(0,1)$; and $\frac{\mu_1\hat{D}_2}{\mu_2\hat{D}_1}x_1^2$ is increasing in $(0,1)$, and 
$f_1(x^{NS}_1)=\frac{\mu_1\hat{D}_2}{\mu_2\hat{D}_1}(x^{NS}_1)^2$. Then we can conclude that $f_2(x_1^{NS})<f_1(x_1^{NS})=\frac{\mu_1\hat{D}_2}{\mu_2\hat{D}_1}(x_1^{NS})^2$ and $f_2(x_1^{SI})>f_1(x_1^{SI})$. This indicates that $f_2(x_1)=f_1(x_1)$ has a unique positive root $x_1^R\in (x^{NS}_1, x^{SI}_1)\subset (0,1)$ based on  Theorem  \ref{th2:allocation}.\\

Similarly, we can show that $x_1^{SI}<x_1^R<x_1^{NS}$ when $\frac{\mu_1}{\mu_2}<\frac{\hat{D}_2}{\hat{D}_1}$ holds. Therefore, the statement holds.

\end{proof}

\subsection*{Proof of Theorem \ref{th5:allocation}\\}
\begin{proof}
Most of statements in Theorem \ref{th5:allocation} can apply similar arguments in the proof of Theorem \ref{th2:allocation}. Thus, we omit the details except  item 2 and 3.\\

Recall that $f_1(x_1)=(1-x_1)^2$. Let  $rd=\frac{\hat{D}_2}{\hat{D}_1}$ and
$$g_2(x_1)=\frac{\beta_{1,2}rd}{\mu_2}x_1^2+\frac{x_1(1-x_1)}{\mu_2}\left[\left( 1+rd\right)x_1-1+\beta_{1,2}-\mu_2 rd\right]=-\frac{(1+rd)x_1}{\mu_2}\left[ x_1^2-a_1x_1+a_0\right]$$ where $a_1=1+\frac{rd(\mu_2+\beta_{1,2})+1-\beta_{1,2}}{1+rd}$ and $a_0=\frac{1+rd\mu_2-\beta_{1,2}}{1+rd}$. Then $f_1-g_2$ has a unique positive solution $x_1^*\in(0,1)$ by applying the argument in Theorem \ref{th2:allocation}.\\

If $1+rd\mu_2-\beta_{1,2}>0$, then $a_0>0$ which implies that $g(x_1)\approx -a_0x_1<0$ when $x_1$ is small. Notice that $g_2(x_1)$ is a degree three polynomial with its coefficient of the degree being negative, $g_2(0)=0$ and $g_2(1)>0$, then we can deduce that  $g_2(x_1)$ has zero and two positive number $\hat{x}_i,i=1,2$ where $\hat{x}_2>1$ and $\hat{x}_1=\frac{a_1-\sqrt{a_1^2-4a_0}}{2}\in (0,1)$. This indicates that 
$$g_2(x_1)<0 \mbox{ if } 0<x_1<\hat{x}_1 \mbox{ and } g_2(x_1)>0 \mbox{ if } \hat{x}_1 <x_1<1.$$
Since $f_1(x_1)=(1-x_1)^2$ is strictly positive and decreasing from 1 to 0 when $x_1$ is increasing from 0 to 1, thus the unique solution $x_1^*>\hat{x}_1$ when $\beta_{1,2}<1+rd\mu_2$.\\


If $1+rd\mu_2-\beta_{1,2}<0$, i.e., $\beta_{1,2}>1+rd\mu_2$, then $g_2(x_1)$ has a unique positive root $\hat{x}_2=\frac{a_1+\sqrt{a_1^2-4a_0}}{2}$ and a unique negative root $\hat{x}_1=\frac{a_1-\sqrt{a_1^2-4a_0}}{2}$where $ g_2(x_1)>0 \mbox{ if } 0<x_1<\hat{x}_2 .$
Recall that $g_2(x_1)$ is a degree three polynomial with its coefficient of the degree being negative, $g_2(0)=0$ and $g_2(1)>0$, then we can conclude that $\hat{x}_2>1$, thus $g_2(x_1)>0$ for all $x_1\in (0,1)$ when $\beta_{1,2}>1+rd\mu_2$. \\

In addition,  the function $g_2(x_1)$ is an increasing function with respect to $\beta_{1,2}$ since
$$\frac{\partial g_2(x_1)}{\partial \beta_{1,2}}=\frac{x_1(rd+(1-x_1))}{\mu_2}>0.$$
If $\beta_{1,2}>1+rd\mu_2$, then the discussion above implies that $g_2(x_1)>0$ for all $x_1\in (0,1)$. Notice that
$$\frac{\partial g_2(x_1)}{\partial \mu_2}=-\frac{g_2(x_1)}{\mu_2^2}-\frac{rd\beta_{1,2}x_1^2}{\mu_2}<0,$$
therefore $g_2(x_1)$ is decreasing with respect to $\mu_2$ when $\beta_{1,2}>1+rd\mu_2$ holds.\\

Since $f_1(x_1)$ is decreasing in $x_1$, the argument above implies that increasing the value of  $\beta_{1,2}$, the intercept of $g_2(x_1)=f_1(x_1)$ is decreasing, i.e., the unique positive solution $x^*_1$ of $g_2(x_1)=f_1(x_1)$ is decreasing with respect to $\beta_{1,2}$.
 Similarly, we can conclude that the unique positive solution $x^*_1$ of $g_2(x_1)=f_1(x_1)$ is increasing with respect to $\mu_2$ when $\beta_{1,2}>1+rd\mu_2$ holds. This also implies that
$$\frac{d N^*_+}{d \mu_1}=\frac{\partial N^*_+}{\partial \mu}\frac{\partial \mu}{\partial \mu_1}+\frac{\partial N^*_+}{\partial \mu}\frac{\partial \mu}{\partial x^*_1}\frac{\partial x_1^*}{\partial \mu_1}<0,$$
$$\frac{\partial N^*_+}{\partial \beta_{1,2}}=\frac{\partial N^*_+}{\partial \mu}\frac{\partial \mu}{\partial x^*_1}\frac{\partial x_1^*}{\partial \beta_{1,2}}<0.$$
Therefore, the total population $N^*_+$ is a decreasing function of $\beta_{1,2}$ and $\mu_1$, respectively. Hence, the statements hold.\\

\end{proof}

\section*{Acknowledgements}
Y.K's research is partially supported by NSF-DMS (Award Number 1313312), Simons Collaboration Grants for Mathematicians (208902). 



\section*{References}

\begin{thebibliography}{10}
\bibitem{Abril2014}S. Abril and C. Gómez, 2014. Strength in numbers: Large and permanent colonies have higher queen oviposition rates in the invasive Argentine ant (Linepithema humile, Mayr). \emph{Journal of Insect Physiology}, \textbf{62},21-25.

\bibitem{Anderson1999}C. Anderson and F. L. W. Ratnieks, 1999. Task partitioning in foraging: effect of colony size on queueing times and information reliability. Pages 31-50 in: Detrain C, Deneubourg J L, Pasteels J M (eds.). \emph{Information Processing in Social Insects}. Birkhauser, Basel. 
\bibitem{Beekman2001}M. Beekman, D. J. T. Sumpter and F. L. W. Ratnieks, 2001. Phase transition between disordered and ordered foraging in Pharaoh's ants. \emph{Proceedings of the National Academy of Sciences}, \textbf{98}(17), 9703-9706.

\bibitem{Beshers2001}S. N. Beshers and J. H. Fewell, 2001. Models of division of labor in social insects, \emph{Annual Review of Entomology}, \textbf{46}, 413-440.
\bibitem{Bonabeau1996}E. Bonabeau, G. Theraulaz, J.-L. Deneubourg, 1996. Quantitative study of the fixed threshold model for the regulation of division of labour in insect societies. \emph{Proceedings of the Royal Society of London. Series B: Biological Sciences}, \textbf{263}(1376), 1565-1569.
\bibitem{Bonabeau1997a}E. Bonabeau, G. Theraulaz, J.-L. Deneubourg, S. Aron and S. Camazine, 1997a. Self-Organization in Social Insects, \emph{Trends in Ecology and Evolution}, \textbf{12}, 188-193.
\bibitem{Bonabeau1997b}E. Bonabeau, A. Sobkowski, G. Theraulaz and  J.-L. Deneubourg, 1997b. Adaptive task allocation inspired by a model of division of labor in social insects, \emph{Proceeding Biocomputing and emergent computation}: Proceedings of BCEC97, 1997, 36-45.
\bibitem{Bonabeau1998a}E. Bonabeau, G. Theraulaz and J.-L. Deneubourg, 1998a. Group and mass recruitment in ant colonies: the influence of contact rates, \emph{Journal of Theoretical Biology}, \textbf{195}, 157-166.

\bibitem{Bonabeau1998b}E. Bonabeau, G. Theraulaz and J.-L. Deneubourg, 1998b. Fixed response thresholds and the regulation of division of labor in insect societies. \emph{Bulletin of Mathematical Biology}, \textbf{60}(4), 753-807.

\bibitem{Bouwma2006}A. M. Bouwma, E. V. Nordheim and R. L. Jeanne, 2006. Per-capita productivity in a social wasp: no evidence for a negative effect of colony size. \emph{Insect. Soc.} \textbf{53}(4), 412-419.
\bibitem{Burd2008}M. Burd and J. J. Howard, 2008. Optimality in a partitioned task performed by social insects, \emph{Biology Letters}, \textbf{4}, 627-629.
\bibitem{Calabi1989}P. Calabi  and J. F. A. Traniello, 1989. Behavioral flexibility in age castes of the ant Pheidole dentata. \emph{J. Insect Behav.} \textbf{2}, 663-677.
\bibitem{Camazine1991}S. Camazine, 1991. Self-organizing pattern formation on the combs of honey bee colonies, \emph{Behavioral Ecology and Sociobiology}, \textbf{28}, 61-76.
 \bibitem{Camazine1999}S. Camazine, P. K. Visscher, J. Finley and R. S. Vetter, 1999. House-hunting by honey bee swarms: collective decisions and individual behaviors, \emph{Insectes Sociaux}, \textbf{46}, 348-360.
\bibitem{Camazine2001}S. Camazine, J.-L. Deneubourg, N. Franks, J. Sneyd, G. Theraulaz and E. Bonabeau, 2001. \emph{Self-Organization in Biological Systems}. Princeton: Princeton University Press.
\bibitem{Camazine2003}S. Camazine, 2003. \emph{Self-organization in biological systems}. Princeton University Press.
\bibitem{Charbonneau2013}D. Charbonneau, B. Blonder and A. Dornhaus, 2013. Social Insects: A model system for network dynamics. \emph{Temporal Networks}.
\bibitem{Charbonneau2015}D. Charbonneau and A. Dornhaus, 2015. Workers `specialized? on inactivity: Behavioral consistency of inactive workers and their role in task allocation. \emph{Behavioral Ecology and Sociobiology}, \textbf{69}(9), 1459-1472.
\bibitem{ChittkaMuller2009}L. Chittka and H. Muller, 2009. Learning, specialization, efficiency and task allocation in social insects. \emph{Commun. Integr. Biol.} \textbf{2}, 151-154.
\bibitem{Clark2014}R. Clark  and J. Fewell, 2014. Transitioning from unstable to stable colony growth in the desert leafcutter ant Acromyrmex versicolor. \emph{Behavioral Ecology and Sociobiology}, \textbf{68}(1), 163-171.

\bibitem{Cornejo2014}A. Cornejo, A. Dornhaus, N. Lynch and R. Nagpal, 2014. Task allocation in ant colonies. \emph{Chapter:
Distributed Computing}, Volume 8784 of the series Lecture Notes in Computer Science, 46-60.
\bibitem{Couzin2005}I. D. Couzin, J. Krause, N. R. Franks and S. A. Levin, 2005. Effective leadership and decision making in animal groups on the move, \emph{Nature}, \textbf{433}, 513-516.

{\bibitem{Cucker1}F. Cucker and S. Smale, 2007a. Emergent behavior in flocks, \emph{IEEE Transactions on Automatic Control}, \textbf{52}, 852.}
{\bibitem{Cucker2}F. Cucker and S. Smale, 2007b. On the mathematics of emergence, \emph{Japanese Journal of Mathematics}, \textbf{2}, 197-227.}
\bibitem{Deneubourg1987}J.-L. Deneubourg, S. Goss, J. M. Pasteels, D. Fresneau and J. P. Lachaud, 1987. Self-organization mechanisms in ant societies II. Learning in foraging and division of labour, \emph{Experientia Supplementum}, \textbf{54}, 177-196.
\bibitem{Deneubourg1989}J.-L. Deneubourg and S. Goss, 1989. Collective patterns and decision-making, \emph{Ethology Ecology \& Evolution}, \textbf{1}, 295-311.
\bibitem{Detrain2006}C. Detrain and J.-L. Deneubourg, 2006. Self-organized structures in a superorganism: do ants ``behave" like molecules? \emph{Physics of Life Reviews}, \textbf{3}, 162-187.
\bibitem{Detrain2008}C. Detrain and J.-L. Deneubourg, 2008. Collective decision-making and foraging patterns in ants and honeybees, \emph{Advances in Insect Physiology}, \textbf{35}, 123-173.
\bibitem{Dorigo1996}M. Dorigo, V. Maniezzo and A. Colorni, 1996.  Ant system: optimization by a colony of cooperating agents, \emph{IEEE Transactions on Systems, Man, and Cybernetics-Part B}, \textbf{26}, 29-41.
\bibitem{Dornhaus2008a}A. Dornhaus, 2008a. Specialization does not predict individual efficiency in an ant. \emph{PLoS Biology}, \textbf{6}(11), e285.
\bibitem{Dornhaus2008b}A. Dornhaus, J-A. Holley, V. G. Pook, G. Worswick and N. R. Franks, 2008b. Why do not all workers work? Colony size and workload during emigrations in the ant Temnothorax albipennis. \emph{Behavioral Ecology and Sociobiology}, \textbf{63}(1), 43-51.
\bibitem{Duarte2011}A. Duarte, F. J. Weissing, I. Pen and L. Keller, 2011. An evolutionary perspective on self-organized division of labor in social insects. \emph{Annual Review of Ecology, Evolution, and Systematics}, \textbf{42}, 91-110.

\bibitem{Dolezal2013}A. G. Dolezal, J. Johnson, B. Holldobler and G. V. Amdam, 2013. Division of labor is associated with age-independent changes in ovarian activity in Pogonomyrmex californicus harvester ants. \emph{Journal of Insect Physiology}, \textbf{59}(4), 519-524.
\bibitem{Eberl2010}H. J. Eberl, M. R. Frederick and P. G. Kevan, 2010. Importance of brood maintenance terms in simple models of the honeybee - varroa destructor - acute bee paralysis virus complex. \emph{Electronic Journal of Differential Equations}, \textbf{19}, 85-98.
{\bibitem{Eftimie}R. Eftimie, 2012. Hyperbolic and kinetic models for self-organized biological aggregations and movement: a brief review,
\emph{Journal of Mathematical Biology}, \textbf{65}, 35-75.}
\bibitem{Elgar1989}M. A. Elgar,1989. Predator vigilance and group size in mammals and birds: a critical review of the empirical evidence. \emph{Biological Reviews}, \textbf{64},13-33.

\bibitem{Fewell1992}J. H. Fewell and M. L. Winston, 1992. Colony state and regulation of pollen foraging in the honey
bee, \emph{Apis mellifera L}, \emph{Behavioral Ecology and Sociobiology}, \textbf{30}, 387-393.
\bibitem{Fewell1999}J. H. Fewell and R. E. Jr Page, 1999. Emergence of division of labour in forced associations of normally solitary ant queens. \emph{Evolutionary Ecology Research}, \textbf{1}, 537-548.
\bibitem{Fewell2003}J. H. Fewell, 2003. Social Insect Networks, \emph{Science}, \textbf{301}, 1867-1870.


\bibitem{Fewell2009}J. H. Fewell, S. Schmidt and T. Taylor, 2009. Division of labor in the context of complexity, Book chapter of \emph{Organization of Insect Societies: From Genomes to Socio-complexity}, Harvard University Press.

\bibitem{Franks2009}N. R. Franks, A. Dornhaus, J. A. R. Marshall and F.-X. D. Moncharmont, 2009. The dawn of a golden age in mathematical insect sociobiology. In: J. Gadau and J. H. Fewell (Eds.), \emph{Organization of Insect Societies: From Genome to Socio-complexity}, Harvard University Press, pp. 437-459.
\bibitem{Gadau2009}J. Gadau and J. H. Fewell, 2009. \emph{Organization of Insect Societies: From Genome to Socio-complexity}. Harvard University Press, Cambridge, MA.
\bibitem{Gautrais2002}J. Gautrais, G. Theraulaz, J.-L. Deneubourg and C. Anderson, 2002. Emergent polyethism as a consequence of increased colony size in insect societies. \emph{Journal of Theoretical Biology}, \textbf{215}, 363-373.
\bibitem{Giraldo2014}Y. Giraldo and J. A. Traniello, 2014. Worker senescence and the sociobiology of aging in ants. \emph{Behavioral Ecology and Sociobiology}, \textbf{68}(12), 1901-1919.

\bibitem{Giray2000}T. Giray, E. Guzmán-Novoa, C. W. Aron, B. Zelinsky, S. E. Fahrbach and G. E. Robinson, 2000. Genetic variation in worker temporal polyethism and colony defensiveness in the honey bee, Apis mellifera. \emph{Behavioral Ecology}, \textbf{11}(1), 44-55.
\bibitem{Gordon1996}D. M. Gordon, 1996. The organization of work in social insect colonies, \emph{Nature}, \textbf{380}, 121-124.
\bibitem{Gordon1999a}D. M. Gordon DM,1999. \emph{Interaction patterns and task allocation in ant colonies}. Basel, Switzerland: Birkhäuser Verlag.
\bibitem{Gordon1999b}D. M. Gordon and N. J. Mehdiabadi, 1999. Encounter rate and task allocation in harvester ants. \emph{Behavioral Ecology and Sociobiology}, \textbf{45}, 370-377.
\bibitem{Gordon2003}D. M., Gordon, 2003. The organization of work in social insect colonies, \emph{Complexity}, \textbf{8}, 43-46.

\bibitem{Gordon2010}D. M. Gordon, 2010.  \emph{Ant Encounters: Interaction Networks and Colony Behavior}. Princeton University Press.
\bibitem{Gordon2011}D. M. Gordon, 2011. The fusion of ecology and behavioral ecology, \emph{Behavioral Ecology}, \textbf{22}, 225-230.
\bibitem{Hamann2013}H. Hamann, I. Karsai and T. Schmickl, 2013. Time delay implies cost on task switching: a model to investigate the efficiency of task partitioning. \emph{Bulletin of Mathematical Biology}, \textbf{75},1181- 1206.
\bibitem{Hee2000}J. J. Hee, D. A. Holway, A. V. Suarez and T. J. Case, 2000. Role of propagule size in the success of incipient colonies of the invasive argentine ant. \emph{Conservation Biology}, \textbf{14}(2), 559-563.
\bibitem{Hou2010}C. Hou, M. Kaspari, H. B. V. Zanden and J. F. Gillooly, 2010. Energetic basis of colonial living in social insects. \emph{Proceedings of the National Academy of Sciences of the United States of America}, \textbf{107}(8), 3634-3638.


\bibitem{Holbrook2009}C. T. Holbrook, R. M. Clark, R. Jeanson, S. M. Bertram, P. F. Kukuk and J. H. Fewell, 2009. Emergence and consequences of division of labor in associations of normally solitary sweat bees, \emph{Ethology}, \textbf{115}, 301-310.
\bibitem{Holbrook2011}C. T. Holbrook, P. Barden and J. H. Fewell, 2011.  Division of labor increases with colony size in the ant, \emph{Pogonomyrmex californicus},  \emph{Behavioral Ecology}, \textbf{22}, 960-966.
\bibitem{Holbrook2013a}C. T. Holbrook,  P. F. Kukuk and J. H. Fewell, 2013a.  Increased group size promotes task specialization in a normally solitary halictine bee. \emph{Behavior}, \textbf{150}, 1449-1466.
\bibitem{Holbrook2013b}C. T. Holbrook, T. H. Eriksson, R. P. Overson, J. Gadau and J. H. Fewell, 2013b. Colony-size effects on task organization in the harvester ant Pogonomyrmex californicus. \emph{Insect. Soc.}, \textbf{60}(2), 191-201.

\bibitem{Holldobler1990}B. Holldobler and E.O. Wilson, 1990. \emph{The Ants}. Harvard University Press.
\bibitem{Hou2008}C. Hou, W. Zuo, M. E. Moses, W. H. Woodru, J. H. Brown and G. B. West, 2008. Energy Uptake and Allocation During Ontogeny. \emph{Science}, \textbf{322} (5902), 736-739.
\bibitem{Ingram2013}K. K. Ingram, A. Pilko, J. Heer and D. M. Gordon, 2013. Colony life history and lifetime reproductive success of red harvester ant colonies. \emph{Journal of Animal Ecology}, \textbf{82}, 540-550.

\bibitem{Jandt2009}J. M. Jandt, E. Huang and A. Dornhaus,  2009. Weak specialization of workers inside a bumble bee nest. \emph{Behavioral Ecology and Sociobiology}, \textbf{63}, 1829-1836.
\bibitem{Jandt2011}J. M. Jandt and A. Dornhaus,  2011. Competition and cooperation: bumblebee spatial organization and division of labor may affect worker reproduction late in life? \emph{Behavioral Ecology and Sociobiology}, \textbf{65}, 2341-2349.
\bibitem{Jandt2012}J. M. Jandt, N. S. Robins, R. E. Moore and A. Dornhaus, 2012. Individual bumblebees vary in response to disturbance: a test of the defensive reserve hypothesis? \emph{Insectes sociaux}, \textbf{59}, 313-321.
\bibitem{Jeanne1996}R. L. Jeanne and E. V. Nordheim, 1996. Productivity in a social wasp: per capita output increases with swarm size. \emph{Behavioral Ecology}, \textbf{7}(1), 43-48.
\bibitem{Jeanson2007}R. Jeanson, J. H. Fewell, R. Gorelick and S. Bertram, 2007. Emergence of increased division of labor as a function of group size, \emph{Behavioral Ecology and Sociobiology}, \textbf{62}, 289-298.
\bibitem{Jeanson2008}R. Jeanson and J. H. Fewell, 2008. Influence of the social context on division of labor in ant foundress associations, \emph{Behavioral Ecology}, \textbf{19}, 567-574.
\bibitem{Julian1999}G. E. Julian and S. Cahan, 1999. Undertaking specialization in the desert leaf-cutter ant Acromyrmex versicolor. \emph{Animal Behaviour}, \textbf{58}, 437-452.


\bibitem{Kang2011-m}Y. Kang, R. Clark, M. Makiyama and J. Fewell, 2011. Mathematical modeling on obligate mutualism: Interactions between leaf-cutter ants and their fungus garden, \emph{Journal of Theoretical Biology}, \textbf{289}, 116-127.
\bibitem{Kang2015}Y. Kang and J. Fewell, 2015. Coevolutionary dynamics of a host-parasite interaction model: obligatory versus facultative parasitism. Accepted in \emph{Natural Resource Modeling}.
\bibitem{Kang2015b}Y. Kang, K. Blanco, T. Davis, Y. Wang and G. DeGrandi-Hoffman, 2015. Disease dynamics of Honeybees with Varroa destructor as parasite and virus vector. Submitted to \emph{Mathematical Biosciences}. Under revision.



\bibitem{Karsai1995}I. Karsai and G. Theraulaz, 1995. Nest building in a social wasp: postures and constraints, \emph{Sociobiology}, \textbf{26}, 83-114.

\bibitem{Karsai1998}I. Karsai and J. W. Wenzel, 1998. Productivity, individual-level and colony-level flexibility and organization of work as consequences of colony size, \emph{Proceedings of the Royal Society B: Biological Sciences}, \textbf{256}, 1261-1268.

\bibitem{Karsai2011}I. Karsai and T. Schmick, 2011. Regulation of task partitioning by a common stomach: a model of nest construction in social wasps. \emph{Behavioral Ecology}, \textbf{22}(4), 819-830.
\bibitem{Karsai2012}I. Karsai and M. D. Phillips, 2012. Regulation of task differentiation in wasp societies: A bottom-up model
of the ``common stomach", \emph{Journal of Theoretical Biology}, \textbf{294}, 98-113.

\bibitem{Keller2009}L. Keller, 2009. Adaptation and the genetics of social behaviour, \emph{Philosophical Transactions of the Royal Society B: Biological Sciences}, \textbf{364}, 3209-3216.
\bibitem{Kerhoas2014}D. Kerhoas, D. Perwitasari-Farajallah, M. Agil, A. Widdig  and A. Engelhardt, 2014. Social and ecological factors influencing offspring survival in wild macaques. \emph{Behavioral Ecology}, \textbf{25},1164-1172.

\bibitem{Kwapich2013}C. L. Kwapich and W. R. Tschinkel, 2013. Demography, demand, death, and the seasonal allocation
of labor in the Florida harvester ant (Pogonomyrmex badius). \emph{Behav Ecol Sociobiol}, \textbf{67}, 2011-2027.

\bibitem{Lanan2012}M. C. Lanan, A. Dornhaus, E. I. Jones, A. Waser and J. L. Bronstein, 2012. The trail less traveled: individual decision-making and its effect on group behavior. \emph{PloS One}, \textbf{7}(10), e47976.
\bibitem{Linksvayer2011}T. Linksvayer, J. H. Fewell, J. Gadau and M. Laubichler, 2011. Developmental evolution in social insects: Regulatory networks from genes to societies, \emph{Journal of Experimental Zoology-Part B: Molecular and Developmental Evolution}.
\bibitem{Marshall2009}J. A. R. Marshall, R. Bogacz, A. Dornhaus, R. Planque,T. Kovacs and N. R. Franks, 2009. On optimal decision-making in brains and social insect colonies. \emph{J. R. Soc. Interface}, \textbf{6}, 1065-1074.
\bibitem{Marsden1976}J. E. Marsden and M. McCracken, 1976. \emph{The Hopf Bifurcation and Its Applications}. Applied
Mathematical Sciences, \textbf{19}, Springer-Verlag , New York, NY.

\bibitem{MaynardSmith1995}J. Maynard Smith and E. Szathm\'ary, 1995. \emph{The Major Transitions in Evolution}. Oxford, England: Oxford University Press.



 \bibitem{Millor1999}J. Millor, M. Pham-Delegue, J.-L. Deneubourg and S. Camazines, 1999. Self-organized defensive behavior in honeybees, \emph{Proceedings of the National Academy of Sciences-USA}, \textbf{96}, 12611-12615.


{\bibitem{Motsch}S. Motsch and E. Tadmor, 2011. A new model for self-organized dynamics and its flocking behavior, \emph{Journal of Statistical Physics}, \textbf{144}, 923 - 947.}

\bibitem{Muscedere2009}M. L. Muscedere, T. A. Willey and J. F. A. Traniello, 2009. Age and task efficiency in the ant Pheidole dentata: young minor workers are not specialist nurses. \emph{Animal Behavior}, \textbf{77}(4), 911-918.
\bibitem{Nicolis2008}S.C. Nicolis, A. Dussutour and E. Despland, 2008. Collective decision-making and behavioral polymorphism in group living organisms,
\emph{Journal of Theoretical Biology}, \textbf{ 254}, 580-586.

\bibitem{Myerscough2004}M. R. Myerscough and B. P. Oldroyd, 2004. Simulation models of the role of genetic variability in social insect task allocation. \emph{Insectes Sociaux}, \textbf{51}(2), 146-152.

\bibitem{Naug1998}D. Naug and R. Gadagkar, 1998. The role of age in temporal polyethism in a primitively eusocial wasp. \emph{Behavioral Ecology and Sociobiology}, \textbf{42}(1), 37-47.

 \bibitem{ODonnell1996}S. O'Donnell, 1996. RAPD markers suggest genotypic effects on forager specialization in a eusocial wasp, \emph{Behavioral Ecology and Sociobiology}, \textbf{38}, 83-88.

\bibitem{ODonnell2007}S. O'Donnell and S. J. Bulova, 2007. Worker connectivity: a review of the design of worker communication systems and their effects on task performance in insect societies. \emph{Insect. Soc.} \textbf{54}(3), 203-210.
\bibitem{Oster1978}G. F. Oster and E. O. Wilson, 1978. \emph{Caste and Ecology in the Social Insects}. Princeton, NJ: Princeton Univ. Press
\bibitem{Pacala1996}S. W. Pacala, D. Gordon and H. C. J. Godfray, 1996. Effects of social group size on information transfer and task allocation, \emph{Evol. Ecol.},\textbf{10}, 127-165.


\bibitem{Page1990}R. E. Page Jr. and S. D. Mitchell, 1990. Self-organization and adaptation in insect societies. Phil. Sci. Assoc.\emph{Fine A., Forbes M. and Wessels L., eds}, \textbf{2}, 289-298.
\bibitem{Page1998}R. E. Page Jr. and S. D. Mitchell, 1998. Self-organization and the evolution of division of labor. \emph{Apidologie}, \textbf{29}(1-2), 171-190.
\bibitem{Page2002}R. E. Page Jr. and J. Erber, 2002. Levels of behavioral organization and the evolution of division of labor, \emph{Naturwissenschaften}, \textbf{89}, 91-106.
\bibitem{Perko2006}L. Perko, 2006. \emph{Differential Equations and Dynamical Systems}. Springer, 3rd Edition.
\bibitem{Pinter-Wollman2011}N. Pinter-Wollman, R. Wollman, A. Guetz, S. Holmes and D. M. Gordon, 2011. The effect of individual variation on the structure and function of interaction networks in harvester ants, \emph{Journal of the Royal Society Interface}, \textbf{8}, 1562-1573.
\bibitem{Pinter-Wollman2012}N. Pinter-Wollman, J. Hubler, J-A. Holley J-A, N. R. Franks and A. Dornhaus, 2012. How is activity distributed among and within tasks in Temnothorax ants? \emph{Behavioral Ecology and Sociobiology}, \textbf{66}, 1407-1420.
\bibitem{Porter1985}S. D. Porter and W. Tschinkel, 1985. Fire ant polymorphism: the ergonomics of brood production. \emph{Behavioral Ecology and Sociobiology}, \textbf{16}(4), 323-336.
\bibitem{Porter1986}S. D. Porter and W. Tschinkel, 1986. Adaptive value of nanitic workers in newly founded red imported fire ant colonies (Hymenoptera: Formicidae). \emph{Annals of the Entomological Society of America}, \textbf{79}, 723-726.
\bibitem{Pratt2009}S. C. Pratt, 2009. Insect societies as model for collective decision making. In: J. Gadau and J. H. Fewell (Eds.), \emph{Organization of Insect Societies: From Genome to Sociocomplexity}, Harvard University Press, 503-524.
\bibitem{Ratti2012}V. Ratti, P. G. Kevan and H. J. Eberl, 2012. A mathematical model for population dynamics in honeybee colonies infested with Varroa destructor and the acute bee paralysis virus. \emph{Canadian Applied Mathematics Quarterly}.
\bibitem{Ravary2007}F. Ravary, E. Lecoutey, G. Kaminski, N. Châline and P. Jaisson, 2007. Individual experience alone can generate lasting division of labor in ants. \emph{Current Biology}, \textbf{17}(15), 1308-1312.
\bibitem{Robinson1989}G. E. Robinson and R. E. Page, 1989. Genetic basis for division of labor in an insect society. \emph{The Genetics of Social Evolution}, eds Breed MD \& Page RE (Westview Press, Boulder, CO), Vol 61-80.
\bibitem{Robinson1992}G. E. Robinson, 1992. Regulation of division of labor in insect societies. \emph{Annual Review of Entomology}, \textbf{37}(1), 637-665.
\bibitem{Robinson2005}G. E. Robinson, C. M. Grozinger and C. W. Whitfield, 2005. Sociogenomics: social life in molecular terms. \emph{Nature Reviews Genetics}, \textbf{6}(4), 257-270.

\bibitem{Seeley1982}T. D. Seeley, 1982. Adaptive significance of the age polyethism schedule in honeybee colonies. \emph{Behavioral Ecology and Sociobiology}, \textbf{11}(4), 287-293.
\bibitem{Seeley1991}T. D. Seeley, S. Camazine and J. Sneyd, 1991.  Collective decision making in honey bees: how colonies
choose among nectar sources. \emph{Behav Ecol Sociobiol}, \textbf{28}, 277-290.
\bibitem{Seeley2009}T. D. Seeley, 2009. \emph{The Wisdom of the Hive: The Social Physiology of Honey Bee Colonies.} Harvard University Press, Cambridge, MA.
\bibitem{Seid2006}M. A. Seid and J. F. A. Traniello, 2006. Age-related repertoire expansion and division of labor in Pheidole dentata (Hymenoptera: Formicidae): a new perspective on temporal polyethism and behavioral plasticity in ants. \emph{Behavioral Ecology and Sociobiology}, \textbf{60}(5), 631-644.

\bibitem{Sendova-Franks1995}A. B. Sendova-Franks and N. R. Franks, 1995. Spatial relationships within nests of the ant Leptothorax unifasciatus and their implications for the division of labour. \emph{Animal Behaviour}, \textbf{50}(1), 121-136.
\bibitem{Smith2006}C. R. Smith and W. R. Tschinkel, 2006 The sociometry and sociogenesis of reproduction in the Florida harvester ant,
Pogonomyrmex badius. \emph{J Insect Sci.}, \textbf{6}, 1-11.
\bibitem{Sumpter2003}D.T. Sumpter and S. Pratt, 2003. A modelling framework for understanding social insect foraging. \emph{Behavioral Ecology and Sociobiology}, \textbf{53}(3), 131-144.
\bibitem{Sumpter2010}D.T. Sumpter, 2010. \emph{Collective Animal Behavior}. Princeton University Press, Princeton, NJ.
\bibitem{Theraulaz1998}G. Theraulaz, E. Bonabeau and J-N. Denuebourg, 1998. Response threshold reinforcements and division of labour in insect societies, \emph{Proceedings of the Royal Society of London B: Biological Sciences}, \textbf{265} (1393), 327-332.

\bibitem{thieme2003mathematics}H. R. Thieme, 2003. \emph{Mathematics in population biology}. Princeton University Press, 2003.


\bibitem{Tofts1992}C. Tofts and N. R. Franks, 1992. Doing the right thing: ants, honeybees and naked molerats. \emph{Trends in Ecology \& Evolution}, \textbf{7}(10), 346-349.
\bibitem{Toth2005}A. L. Toth and G. E. Robinson, 2005. Worker nutrition and division of labour in honeybees. \emph{Animal Behaviour}, \textbf{69}(2), 427-435.

\bibitem{Tschinkel1988}W. R. Tschinkel, 1988. Colony growth and the ontogeny of worker polymorphism in the fire ant, Solenopsis invicta. \emph{Behavioral Ecology and Sociobiology}, \textbf{22},103-115.
\bibitem{Tschinkel1993}W. R. Tschinkel, 1993. Sociometry and sociogenesis of colonies of the fire ant Solenopsis invicta during one annual cycle. \emph{Ecological Mongraphs}, \textbf{63}, 425-427.
\bibitem{Tschinkel1999}W. R. Tschinkel, 1999. Sociometry and sociogenesis of colony-level attributes of the Florida harvester ant (Hymenoptera: Formicidae). \emph{Annals of the Entomological Society of America}, \textbf{92}, 80-89
\bibitem{Wahl2002}L. M. Wahl, 2002. Evolving the division of labour: generalists, specialists and task allocation. \emph{J. Theor. Biol.}, \textbf{219}(3), 371-388.
\bibitem{Udiani2015}O. Udiani, N.Pinter-Wollman and Y. Kang, 2015. Identifying robustness in the regulation of foraging of ant colonies using an interaction based model with backward bifurcation. \emph{Journal of Theoretical Biology}, \textbf{365}, 61-75.
\bibitem{Waibel2006}M. Waibel, D. Floreano, S. Magnenat and L. Keller, 2006. Division of labour and colony efficiency in social insects: effects of interactions between genetic architecture, colony kin structure and rate of perturbations. \emph{Proc. R. Soc. B},  \textbf{273}, 1815-1823.


\bibitem{Wakano1998}J. Y. K. Wakano, K. Nakata and N. Yamamura, 1998. Dynamic model of optimal age polyethism in social insects under stable and fluctuating environments. \emph{J. Theor. Biol.}, \textbf{193}, 153-165

\bibitem{Waters2012}J. S. Waters and J. H. Fewell, 2012. Information processing in social insect networks, \emph{PLoS ONE }, \textbf{7}(7): e40337. DOI:10.1371/journal.pone.0040337.

\bibitem{Watmough1995}J. Watmough and S. Camazine, 1995. Self-organized thermoregulation of honeybee clusters, \emph{Journal of Theoretical Biology}, \textbf{176}, 391-402.
\bibitem{Weidenmuller2004}A. Weidenmuller, 2004. The control of nest climate in bumblebee (Bombus terrestris) colonies: interindividual variability and self reinforcement in fanning response. \emph{Behavioral Ecology}, \textbf{15}(1), 120-128.
\bibitem{Weier1995}J. A. Weier, D. H. Feener Jr and J. R. B. Lighton, 1995. Inter-individual variation in energy cost of running and loading in the seed-harvester ant, Pogonomyrmex maricopa. \emph{Journal of Insect Physiology}, \textbf{41}(4), 321-327.
\bibitem{Wilson1968}E. O. Wilson, 1968. The ergonomics of caste in the social insects, \emph{The American Naturalist}, \textbf{102}, 41-66. 

\bibitem{Wilson1971}E. O. Wilson, 1971. \emph{The Insect Societies.} Cambridge, MA: Harvard Univ. Press 

\bibitem{Wilson1976}E. O. Wilson, 1976. Behavioral discretization and number of castes in an ant species, \emph{Behavioral Ecology and Sociobiology}, \textbf{1}, 141-154. 
\bibitem{Wilson1980a}E. O. Wilson, 1980a. Caste and division of labor in leaf-cutter ants (Hymenoptera :  Formicidae-atta)  I.  The overall pattern in A. Soxdens. \emph{Behavioral Ecology and Sociobiology}, \textbf{7},143-156.
\bibitem{Wilson1980b}E. O. Wilson, 1980b. Caste and division of labor in leaf-cutter ants (Hyman: Formicidae: AHa )  II.  The ergonomics optimization of leaf-cutting. \emph{Behavioral Ecology and Sociobiology}, \textbf{7},157-165.
 \bibitem{Wilson1985a}E. O. Wilson, 1985a. The principles of caste evolution. In \emph{Experimental Behavioral Ecology and Sociobiology}, ed. B. Holldobler and M. Lindauer, pp. 307-324. Stuttgart: Gustav Fischer Verlag 

 \bibitem{Wilson1985b}E. O. Wilson, 1985b. The sociogenesis of insect colonies, \emph{Science}, \textbf{228}, 1489-1495. 

 \bibitem{Wilson1987}E. O. Wilson, 1987. Causes of ecological success-the case of the ants: the 6th Tansley lecture, \emph{Journal of Animal Ecology}, \textbf{56}, 1-9.

  \bibitem{Wilson2009}E. O.  Wilson and Bolldobler, 2009. \emph{The Superorganism}. New York, NY: WW Norton \& Co.
\end{thebibliography}
\end{document}